\documentclass[12pt,preprint2]{emulateapj}
\usepackage{pslatex}
\usepackage[T1]{fontenc}
\usepackage[latin1]{inputenc}
\usepackage{subfigure}
\usepackage{amsmath}
\usepackage{natbib}
\usepackage{graphicx,graphics}
\usepackage{amssymb}
\usepackage[english]{babel}

\def\mh{M_{\rm BH}}
\def\mb{M_{\rm b}}
\def\vk{V_{\rm k}}

\def\rk{r_{\rm k}}
\def\rgc{HCSS}
\def\rgcs{HCSSs}
\def\rh{r_{\bullet}}

\def\rinfl{r_{\rm infl}}
\def\kms{km s$^{-1}$}

\newcommand{\gap}{\;\rlap{\lower 2.5pt \hbox{$\sim$}}\raise 1.5pt\hbox{$>$}\;}
\newcommand{\lap}{\;\rlap{\lower 2.5pt \hbox{$\sim$}}\raise 1.5pt\hbox{$<$}\;}
\newcommand{\beq}{\begin{equation}}
\newcommand{\eeq}{\end{equation}}
\newcommand{\msun}{M_\odot}

\shorttitle{Hypercompact Stellar Systems}
\shortauthors{Merritt, Schnittman \& Komossa}

\begin{document}

\title{Hypercompact Stellar Systems Around Recoiling Supermassive Black Holes} 

\author{David Merritt}
\affil{Center for Computational Relativity and Gravitation and
Department of Physics, Rochester Institute of Technology,
Rochester, NY 14623, USA}
\author{Jeremy D. Schnittman}
\affil{Department of Physics and Astronomy, The Johns Hopkins University,
Baltimore, MD 21218, USA} 
\author{S. Komossa} 
\affil{Max-Planck-Institut f\"ur extraterrestrische Physik,
Postfach 1312, 85741 Garching, Germany}

\begin{abstract}
A supermassive black hole ejected from the center of a galaxy
by gravitational wave recoil carries a retinue of bound stars --
a ``hypercompact stellar system'' (\rgc).
The numbers and properties of \rgcs\ contain information
about  the merger histories of galaxies, the late evolution of binary
black holes, and the distribution of gravitational-wave kicks.
We relate  the structural properties (size, mass, density profile) 
of \rgcs\ to the properties of their host galaxies and to the
size of the kick, in two regimes:
collisional ($\mh\lap 10^7\msun$), i.e. short nuclear relaxation times;
and collisionless ($\mh\gap 10^7\msun$), i.e. long nuclear relaxtion times.
\rgcs\ are expected to be similar in size and luminosity to globular clusters
but in extreme cases (large galaxies, kicks just above escape velocity) 
their stellar mass can approach that of ultra-compact dwarf galaxies.
However they differ from all other classes of compact stellar system
in having very high internal velocities.
We show that the kick velocity is encoded in the velocity dispersion
of the bound stars.
Given a large enough sample of \rgcs, the distribution of
gravitational-wave kicks can therefore be empirically determined.
We combine a hierarchical merger algorithm with stellar population models
to compute  the rate of production of \rgcs\ over time and
the probability of observing \rgcs\ in the local universe as a function of
their apparent magnitude, color, size and velocity dispersion,
under two different assumptions about the star formation history
prior to the kick.
We predict that $\sim 10^2$ \rgcs\ should be detectable within 
2 Mpc of the center of the Virgo cluster 
and that many of these should be bright enough that their kick
velocities (i.e. velocity dispersions) could be measured with 
reasonable exposure times.
We discuss other strategies for detecting \rgcs\ 
and speculate on some exotic manifestations.
\end{abstract}

%\keywords{galaxies: active -- galaxies: evolution -- galaxies: X-rays}

\section{Introduction}

A natural place to search for supermassive black holes (SMBHs)
is at the centers of galaxies, where they presumably are born
and spend most of their lives.
But it has become increasingly clear that a SMBH can be violently
separated from its birthplace as a result of linear momentum
imparted by gravitational waves during strong-field 
interactions with other SMBHs \citep{Peres:62,Bekenstein:73,
Redmount:89}.
The largest net recoils are produced from configurations that bring
the two holes close enough together to coalesce.
Kick velocities following coalescence can be as high as 
$\sim 200$ \kms\ in the case of nonspinning holes
\citep{Gonz:Nonspin:07,Sopuerta:07}; 
$\sim 4000$ \kms\ for maximally spinning, equal mass BHs on initially 
circular orbits \citep{Campanelli:07,Gonz:Spin:07,Herrmann:07,Pollney:07,
Tichy:07,Brugmann:08,Dain:08,Baker:08};
and even higher, $\sim 10,000$ \kms, for black holes
that approach on nearly-unbound orbits \citep{Healy:08}.
Since escape velocities from the centers of even the largest galaxies are
$\lap 2000$ \kms\ \citep{Merritt:04},
it follows that the kicks can in principle remove SMBHs completely 
from their host galaxies.
While such extreme events may be relatively rare 
\citep[e.g.][]{SB:07,Schnittman:07}, 
recoils large enough to displace SMBHs at least temporarily from galaxy cores
-- to distances of several hundred to a few thousand parsecs -- may be 
much more common \citep{Merritt:04,MQ:04,GM:08,KM:08b}.

Komossa et al. (2008) reported the detection of a recoil candidate.
This quasar exhibits a kinematically offset broad-line region with
a velocity of 2650 km s$^{-1}$ , and very narrow, restframe, 
high-excitation emission lines which lack the usual ionization 
stratification  -- two key signatures of kicks. 
In addition to spectroscopic signatures \citep{Naked,Bonning:07},
recoiling SMBHs could be detected by their soft X-ray, UV and IR flaring 
\citep{Shields:08,Lippai:08,SK:08}
resulting from shocks in the accretion disk surrounding the coalesced SMBH.
Detection of recoiling SMBHs  in this way
is contingent on the presence of gas.
But only a small fraction of {\it nuclear} SMBHs
exhibit signatures associated with gas accretion, 
and a SMBH that has been displaced
from the center of its galaxy will only shine as
a quasar until its bound gas has been used up \citep{Loeb:07}.
The prospect that the SMBH will encounter and capture
significant amounts of gas on its way out are small
\citep{Kapoor:76}.

A SMBH ejected from the center of a galaxy will always carry
with it a retinue of bound stars. 
The stars can reveal themselves via tidal disruption flares
or via accretion of gas from stellar winds onto the SMBH
\citep[][hereafter Paper I]{KM:08a}. 
The cluster of stars is itself directly observable,
and that is what we discuss in the current work.
The linear extent of such a cluster is fixed by 
the magnitude $\vk$ of the kick velocity and 
by the mass of the SMBH:
\begin{subequations}
\begin{eqnarray}
\rk &\equiv& G\mh/\vk^2 \\
&\approx& 0.043\ {\rm pc} \left({\mh\over 10^7\msun}\right)
\left({\vk\over 10^3\ {\rm km\ s}^{-1}}\right)^{-2}\, .
\end{eqnarray}
\label{eq:rk}
\end{subequations}
Reasonable assumptions about the density of stars
around the binary SMBH prior to the kick (Paper I)
then imply a total luminosity of the bound population 
comparable to that of a globular star cluster.

In this paper we discuss the properties of these ``hyper-compact
stellar systems'' (\rgcs) and their relation to host galaxy
properties.
Our emphasis is on the prospects for detecting such objects
in the nearby universe at optical wavelengths,
and so we focus on the properties
that would distinguish \rgcs~from other stellar
systems of comparable size or luminosity.
As noted in Paper I, a key signature is their high internal 
velocity dispersion:
because the gravitational force that binds the cluster
comes predominantly from the SMBH, of mass 
$10^6\lap M_{\rm BH}/\msun\lap 10^9$,
stellar velocities will be much higher than in ordinary 
stellar systems of comparable luminosity.
Other signatures include the small sizes of \rgcs\
(unfortunately, too small to be resolved except for
the most nearby objects); their high space velocities
(due to the kick);
and their broad-band colors, which should resemble more
closely the colors of galactic nuclei rather than the colors of
uniformly old and metal-poor systems like globular clusters.

As we discuss in more detail below (\S2), a remarkable property
of \rgcs\ is that they encode, via their internal kinematics,
the velocity of the kick that removed them from their host galaxy.
A measurement of the velocity dispersion of the
stars in a \rgc\ is tantamount to a measurement of the
amplitude of the kick -- independent of how long ago
the kick occurred; the black hole mass; and the space velocity
of the \rgc\ at the moment of observation.
This property of \rgcs\ opens the door to an empirical determination
of the distribution of gravitational-wave kicks.
 
The outline of the paper is as follows.
\S 2 derives the relations between the structural parameters
of \rgcs -- mass, radius, and internal velocity dispersion --
given assumed values for the slope and density normalization
of the stellar population around the SMBH just before the kick.
In \S3, models for the evolution of binary SMBHs are reviewed
and their implications for the pre-kick distribution of stars
are described.
These results, combined with the relations derived in \S2, 
allow us to relate the structural parameters of \rgcs~to the
global properties of the galaxies from which they were ejected.
\S4 discusses the effect of post-kick dynamical evolution of
the \rgcs~on their observable properties.
Stellar evolutionary models are used to predict the luminosities 
and colors of \rgcs~and their post-kick evolution in \S5,
and in \S6, the evolutionary models are combined with models
of hierarchical merging to estimate
the number of \rgcs~to be expected per unit volume in the local
universe as a function of their observable properties.
\S7 discusses search strategies for \rgcs\ and various other 
observable signatures that might be uniquely associated with
them. 
In \S8 we briefly discuss the inverse problem of reconstructing
the distribution of recoil velocities from an observed
sample of \rgcs.
\S9 sums up and suggests topics for further investigation.

\section{Structural Relations}

In what follows, we adopt the $\mh-\sigma$ relation in the form given by 
Ferrarese \& Ford (2005):
\beq
{\mh\over 10^8\msun} = 1.66 
\left({\sigma\over 200\ {\rm km\ s}^{-1}}\right)^{4.86}\, ,
\label{eq:MS}
\eeq
with $\sigma$ the 1-D velocity dispersion of the galaxy bulge.
The influence radius of the SMBH is defined as
\begin{subequations}
\begin{eqnarray}
\rinfl &\equiv& {G\mh\over\sigma^2} \\
&\approx& 10.8 {\rm pc} \left({\mh\over 10^8\msun}\right)
\left({\sigma\over 200\ {\rm km\ s}^{-1}}\right)^{-2}
\end{eqnarray}
\label{eq:defrinfl}
\end{subequations}
and $\rk\approx (\sigma/\vk)^2\rinfl$.

\subsection{Bound Population}

As discussed in Paper I,
a recoiling SMBH carries with it a cloud of stars on bound orbits.
Just prior to the kick, most of the stars that will remain
bound lie within a sphere of radius $\sim\rk$ around the
SMBH (eq.~\ref{eq:rk}).
Setting $\mh=3\times 10^6\msun$ and $\vk= 4000$ \kms
gives $\rk\approx 10^{-3}$ pc as an approximate, 
minimum expected value for the size of a \rgc;
such a small size justifies the adjective ``hypercompact''.
The largest values of $\rk$ would probably be associated
with \rgcs\ ejected from the most massive galaxies,
containing SMBHs with masses
$\mh\approx 3\times 10^9\msun$ and travelling with
a velocity just above escape, $\sim 2000$ \kms;
this implies $\rk\approx$ several pc -- similar to
a large globular cluster.

Assuming a power law density profile before the kick,
$\rho(r)=\rho(r_0)(r/r_0)^{-\gamma}$, the stellar mass $M_{\rm k}$
{\it initially} within radius $\rk$ is
\begin{subequations}
\begin{eqnarray}
M_{\rm k} \equiv M(r\le\rk) &=& {4\pi\over 3-\gamma} \rho(\rk)\rk^3 \\
&=& {4\pi\over 3-\gamma} \rho_0 r_0^\gamma 
\left({G\mh\over \vk^2}\right)^{3-\gamma}\, ,
\end{eqnarray}
\label{eq:mk1}
\end{subequations}
where $\rho_0\equiv\rho(r_0)$.
As a fiducial radius at which to normalize the pre-kick density
profile, we take $r_0=\rh$, defined as the radius containing an 
integrated mass in stars equal to twice $\mh$.
(We expect $\rh$ to be of order $\rinfl$; see \S 3 for a further
discussion.)
Equation~(\ref{eq:mk1}) then becomes
\beq
M_{\rm k} = 2\mh \left({G\mh\over \rh\vk^2}\right)^{3-\gamma}.
\eeq
After the kick, the density profile will be nearly
unchanged at $r<\rk$ but will be strongly truncated at
larger radii.
We define $M_{\rm b}$ to be the total mass in stars that remain
bound to the SMBH after the kick, and write
\begin{subequations}
\begin{eqnarray}
f_{\rm b}\equiv {\mb\over\mh} 
&=& F_1(\gamma)\left({G\mh\over\rh\vk^2}\right)^{3-\gamma} \\
&\propto& \vk^{-2(3-\gamma)} \, ,
\end{eqnarray}
\label{eq:f1}
\end{subequations}
where 
\begin{equation}
2\mb = F_1(\gamma)M_{\rm k} .
\label{eq:F_1}
\end{equation}

Kicks large enough to remove a SMBH from a galaxy core
must exceed $\sigma$, 
and escape from the galaxy implies $(\vk/\sigma)^2\gg 1$;
hence $\rk\ll\rinfl$ to a good approximation.
It follows that stars that remain bound following the kick will be 
moving essentially in the point-mass potential of the SMBH both 
before and after the kick.
To the same order of approximation, the SMBH's velocity
is almost unchanged as it climbs out of the galaxy potential well
(at least during the relatively short time required for the stars
to reach a new steady state distribution after the kick).
Finally, since the bulk of the recoil is imparted to the SMBH
in a time $\sim G\mh/c^3$, the kick is essentially instantaneous
as seen by stars at distances $r\gap G\mh/c^2\ll\rk$
\citep{Schnittman:08}.

These three approximations allow the properties of the bound population 
to be computed uniquely given the initial distribution (Paper I).
Transferring to a frame moving with velocity $\mathbf{V}_{\rm k}$ 
after the kick,
the stars respond as if they had received an implusive velocity 
change $-\mathbf{V}_{\rm k}$ at the instant of the kick, causing the
elements of their Keplerian orbits about the SMBH to instantaneously change.
As a result, all initially-bound stars outside of the sphere $r=8\rk$ at the
moment of the kick acquire positive 
energies with respect to  the SMBH and escape.
Some of the stars initially at $\rk\lap r< 8\rk$ escape while
others remain bound.
The stellar distribution at $r\lap\rk$ is almost unchanged.

\begin{figure}
\includegraphics[clip,width=0.45\textwidth]{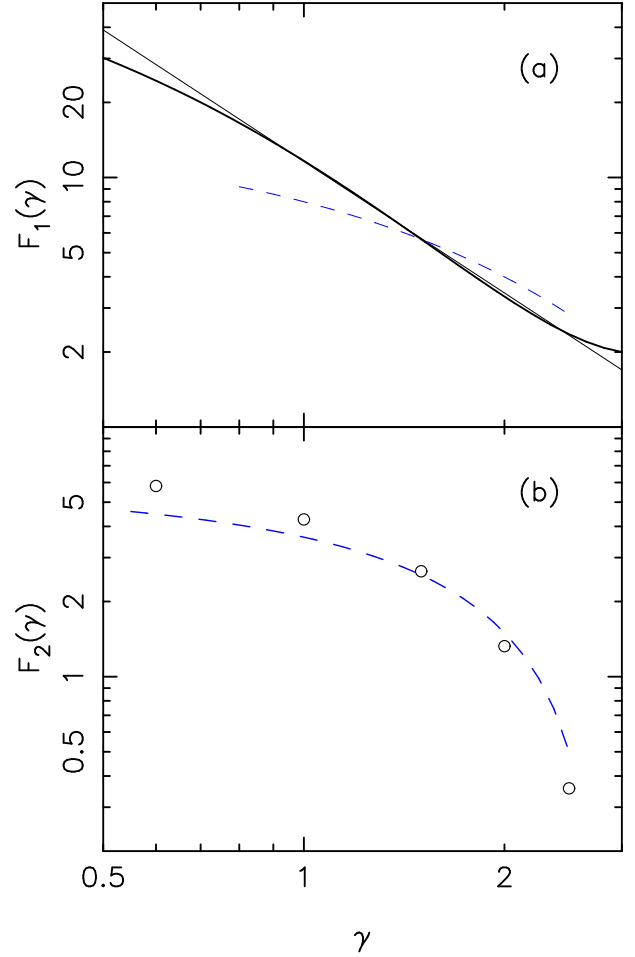}
\caption{Dimensionless factors that describe (a) the
stellar mass bound to a kicked SMBH (eq.~\ref{eq:f1})
and (b) its effective radius (eq.~\ref{eq:f2}).
Thick (black) line in the upper panel is the exact expression
derived in Appendix A and thin (black) line is the approximation,
given in eq.~\ref{eq:fofgamma}.
Open circles in (b) were computed using a Monte-Carlo algorithm.  
Dashed (blue)  lines in both panels show the Dehnen-model approximations
of eqs.~(\ref{eq:md}) and (\ref{eq:reff}). 
}
\label{fig:fgamma}
\end{figure}

\begin{figure}
\includegraphics[clip,width=0.45\textwidth]{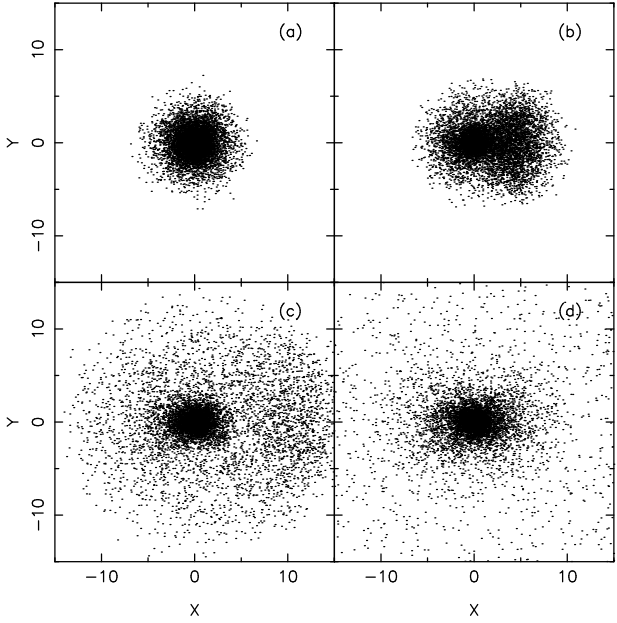}
\caption{Evolution of the  bound population following a kick;
the kick was in the $-\mathbf{X}$ direction at $t=0$.
Each frame is centered on the (moving) SMBH.
Stars were initially distributed as a power law in density,
$\rho\propto r^{-7/4}$; only stars which remain bound
following the kick are plotted.
Unit of length is $\rk$ and frames (a,b,c,d) correspond to 
times of ($0, 10, 30, 100$) in units of $G\mh/\vk^3$.
}
\label{fig:frames}
\end{figure}

Appendix A presents a computation of the bound mass under these
approximations and gives an expression for $F_1(\gamma)$ in
terms of integrals of simple functions.
Figure~\ref{fig:fgamma}a plots this expression, and also the
function
\beq
F_1(\gamma) = 11.6\gamma^{-1.75} \, ,
\label{eq:fofgamma}
\eeq
which is seen to be an excellent approximation for 
$0.7\lap\gamma\lap 2.5$.
(As a check, we computed $F_1(\gamma)$ in another way:
we generated Monte-Carlo samples of positions and velocities
corresponding to an isotropic, power-law distribution of
stars around the SMBH prior to the kick and discarded the
stars that would be unbound after the kick.)
Combining equations~(\ref{eq:f1}) and (\ref{eq:fofgamma}), we get
\beq
f_{\rm b} \approx 11.6 \gamma^{-1.75} 
\left({G\mh\over\rh\vk^2}\right)^{3-\gamma}.
\eeq
Setting $\gamma=1$ in this expression gives
\beq
f_{\rm b} \approx 2\times 10^{-4}
\left({\mh\over 10^7\msun}\right)^{2}
\left({\rh\over 10\ {\rm pc}}\right)^{-2}
\left({\vk\over 10^3\ {\rm km\ s}^{-1}}\right)^{-4},
\label{eq:fb11}
\eeq
which reproduces reasonably well the values for the 
bound mass found by Boylan-Kolchin et al. (2004) in their
$N$-body simulations of kicked SMBHs; their galaxy
models had central power-law density cusps with $\gamma=1$.

Setting $\gamma=1.75$, the value corresponding to a 
collisional (Bahcall-Wolf) cusp, gives
\beq
f_{\rm b} \approx 5\times 10^{-3}
\left({\mh\over 10^7\msun}\right)^{1.25}
\left({\rh\over 10\ {\rm pc}}\right)^{-1.25}
\left({\vk\over 10^3\ {\rm km\ s}^{-1}}\right)^{-2.5},
\label{eq:fb175}
\eeq
which will be useful in what follows.

\begin{figure}
\includegraphics[clip,width=0.45\textwidth]{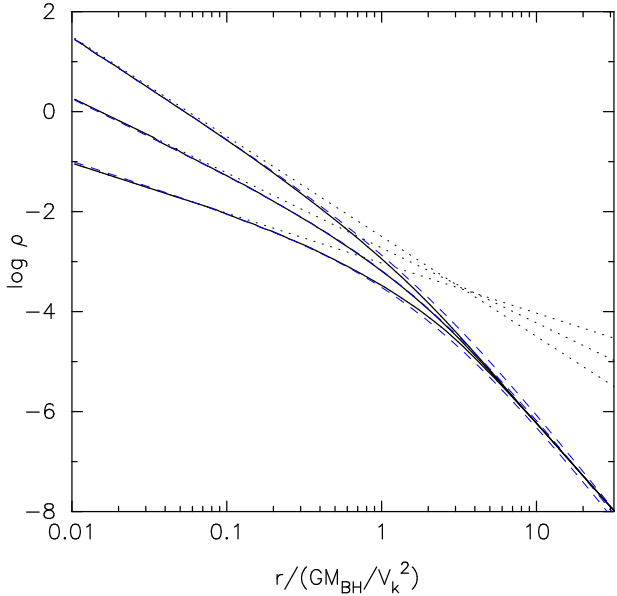}
\caption{Steady state, spherically symmetrized density profiles
of the bound population for $\gamma=(1,1.5,2)$.
Dotted lines show the pre-kick densities;
dashed (blue) lines are Dehnen-model fits.
}
\label{fig:profs}
\end{figure}

Given the elements of the  Keplerian orbits after the kick,
the subsequent evolution of the stellar distribution 
can be computed by simply advancing the positions in time
via Kepler's equation.
(Alternately the stellar trajectories can be brute-force integrated; 
both methods were used as a check.)
Figure~\ref{fig:frames} shows how the bound population evolves
from its initially spherical configuration, into a fan-shaped
structure at $t\approx 10 (G\mh/\vk^3)$, and finally into
a reflection-symmetric, elongated spheroid with major axis in the 
direction of the kick at $t\approx 100 (G\mh/\vk^3)$.
The latter time is
\beq
t_{\rm sym} \sim  3\times 10^{3} {\rm yr}
\left({\mh\over 10^7\msun}\right)
\left({\vk\over 1\times 10^3\ {\rm km\ s}^{-1}}\right)^{-3} \, ,
\label{eq:tsym}
\eeq
during which interval the SMBH would travel a distance
\beq
d_{\rm sym} \sim 3 {\rm pc}
\left({\mh\over 10^7\msun}\right)
\left({\vk\over 10^3\ {\rm km\ s}^{-1}}\right)^{-2}.
\label{eq:dsym}
\eeq
Observing the kick-induced asymmetry would only be possible 
for a short time after the kick;
however the elongation of the bound cloud at $r\gg\rk$
would persist indefinitely.

In general, the galactic nucleus might be elongated {\it before}
the kick, and its major axis will be oriented in some random direction
compared with $\mathbf{V}_{\rm k}$.
Since the stellar distribution at $r\lap\rk$ 
is nearly unaffected by the kick, the generic result will be a bound population
that exhibits a twist in the isophotes at $r\approx \rk$
and a radially-varying ellipticity.

Continuing with the same set of approximations made above, we can 
compute the steady-state distribution of the bound population by
fixing  the post-kick elements of the Keplerian orbits and
randomizing the orbital phases (or equivalently by continuing
the integration of Fig.~\ref{fig:frames} until late times.)
The resultant density profiles
are shown in Figure~\ref{fig:profs} for $\gamma=(1,1.5,2)$.
Beyond a few $\rk$, the spherically-symmetrized density falls 
off as $\sim r^{-4}$;
the stars in this extended envelope move on eccentric orbits
that were created by the kick.

It turns out that Dehnen's (1993) density law:
\beq
\rho(r) = {(3-\gamma)M_{\rm D}\over 4\pi} \xi^{-\gamma}
\left(1 + \xi\right)^{\gamma-4}, \ \ \ \ \xi\equiv r/r_{\rm D}
\label{eq:dehnen}
\eeq
is a good fit to these density
profiles for $1\lap\gamma\lap 2$, 
if $r_{\rm D}$ is set to $2.0\rk$; here $M_{\rm D}$ is the total (stellar) mass.
Figure~\ref{fig:profs} shows the Dehnen-model fits
as dashed lines.
Using the expressions in Dehnen (1993), it is easy to show
that the Dehnen models so normalized satisfy
\beq
{M_{\rm D}\over\mh} = 2^{4-\gamma}
\left({G\mh\over\rh\vk^2}\right)^{3-\gamma} \, ,
\label{eq:md}
\eeq
implying $F_1\approx 2^{4-\gamma}$.
This alternate expression for $F_1$ is plotted as the dashed line in 
Figure~\ref{fig:fgamma}a.
Unless otherwise stated, we will use equation~(\ref{eq:fofgamma}) for
$F_1$ in what follows.

So far we have assumed that stars remaining bound to the SMBH
experience only its point-mass force.
In reality, beyond a radius of order $r_{\rm infl}\approx (\vk/\sigma)^2\rk$,
stars will also feel a significant acceleration from the combined
attraction of the other stars, leading to a tidally
truncated density profile at $r\gg\rk$.
We ignore that complication in what follows.

We note that $\rh$ is determined by the density of stars
{\it just before} the massive binary has coalesced,
and may be substantially different from $r_{\rm infl}$
(eq.~\ref{eq:defrinfl}).
In the next section we discuss predictions for $\rh$
based on a number of models for the evolution of the massive
binary prior to the kick.

Before doing so, we first present the mass-radius and
mass-velocity dispersion relations for the bound population,
expressed in terms of $\rh$ as a free parameter.

\subsection{Mass-Radius Relation}

Combining equations~(\ref{eq:rk}) and~(\ref{eq:f1}), we get
\beq
{\mb\over\mh} = F_1(\gamma)\left({\rk\over\rh}\right)^{3-\gamma}\, .
\label{eq:MR}
\eeq
As a measure of the size of the \rgc, the effective 
radius $r_{\rm eff}$, i.e. the radius containing one-half
of the stellar mass in projection, is preferable to $\rk$.
We define a second form factor $F_2$ such that
\beq
r_{\rm eff} = F_2(\gamma) \rk \, .
\label{eq:f2}
\eeq
Figure~\ref{fig:fgamma}(b) plots $F_2(\gamma)$. 
Also shown by the dashed line is the relation corresponding to the
Dehnen-model approximation described above, for which
\beq
F_2(\gamma)\approx 1.5\left(2^{1/(3-\gamma)} -1\right)^{-1} 
\label{eq:reff}
\eeq
(Dehnen 1993).
The Dehnen model approximation is
reasonably good for all $\gamma$ in the range 
$0.5\le\gamma\le 2.5$ and will be used as the default definition
for $F_2$ in what follows.

Combining equations~(\ref{eq:MR}) and (\ref{eq:reff})
gives the mass-radius ($M_{\rm b}-r_{\rm eff}$) relation for \rgc's, 
in terms of the (yet unspecified) $\rh$:
\begin{subequations}
\begin{eqnarray}
\mb &=& K(\gamma) \mh \rh^{\gamma-3} r_{\rm eff}^{3-\gamma}, \\
K(\gamma) &\equiv& 11.6\gamma^{-1.75}
\left[(3/2)(2^{1/(3-\gamma)}-1)^{-1}\right]^{\gamma-3};
\label{eq:mvsr}
\end{eqnarray}
\end{subequations}
for $\gamma=(1,1.5,2)$, $K=(0.89,1.41,2.32)$.

\subsection{Mass-Velocity Dispersion Relation}

Stars bound to a recoiling SMBH move within the 
point-mass potential of the SMBH, for which the local circular velocity
is $(G\mh/r)^{1/2}$.
The circular velocity at $r=\rk$ is just $\vk$, so the
characteristic (e.g. rms) speed of stars in the bound cloud scales
as $\vk$, motivating us to define a third form factor $F_3$ such that
\beq
\sigma_{\rm obs} = F_3(\gamma)\vk ,
\label{eq:defs}
\eeq 
where $\sigma_{\rm obs}$ is the measured velocity dispersion.
To the extent that $\gamma$ is known, and/or the dependence
of $F_3$ on $\gamma$ is weak, it follows that 
the amplitude of the initial
kick can be empirically determined by measuring the
velocity dispersion of the stars.

\begin{figure}
\includegraphics[clip,width=0.45\textwidth]{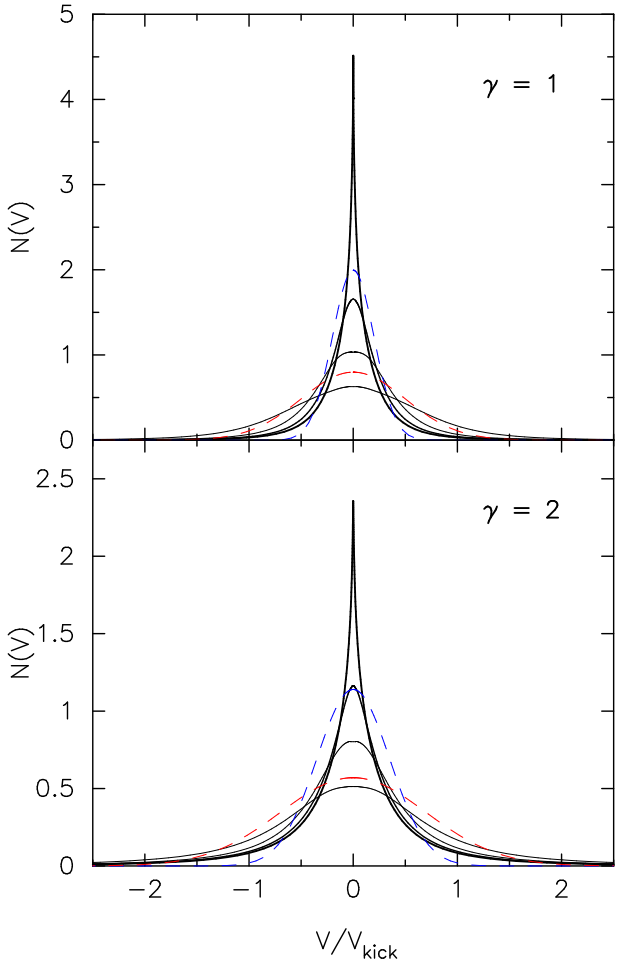}
\caption{Line-of-sight distribution of velocities of stars bound
to a recoiling SMBH, as seen from a direction perpendicular to the kick.
Initially $\rho\propto r^{-\gamma}, \gamma=(1,2)$; 
the phase-space distribution following the kick was computed as in Paper I.
Solid curves show $N(V)$ as defined by all bound stars
(thick) and progressively thinner curves show $N(V)$ defined by 
bound stars within a projected distance of $(10,3,1)\rk$ from the SMBH.
Dashed (blue and red) curves show Gaussian distributions 
with $\sigma=(0.2,0.5)\vk$ ($\gamma=1$) and 
  $\sigma=(0.35,0.75)\vk$ ($\gamma=2$)
respectively.
}
\label{fig:nofv}
\end{figure}

An integrated spectrum will include stars at all (projected) radii
within the spectrograph slit.
(E.g. at the distance of the Virgo cluster, a $1''$ slit corresponds to 
$\sim 80$ pc, larger than $r_{\rm eff}$ for even the largest \rgc's.)
Since $V\sim r^{-1/2}$, the distribution $N(V)$
of line-of-sight velocities of stars within the slit
will contain significant contributions from stars moving both much faster and
much slower than $\vk$ and can be significantly non-Gaussian\footnote{Integrated spectra of the centers of galaxies typically
{\it are} well modelled via Gaussian broadening functions. 
This is because most of the light in the slit comes from stars 
that are far from the SMBH.}.

\begin{figure}
\includegraphics[clip,angle=-90.,width=0.45\textwidth]{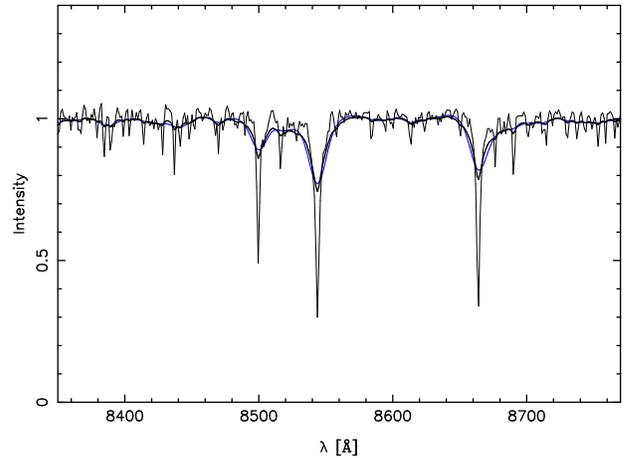}
\caption{Absorption line spectrum of the K0III star HR 7615, 
convolved with two broadening functions.
{\it Thick (black) curve:} $N(V)$ from the top panel of Fig.~\ref{fig:nofv},
computed from the entire bound population, assuming $\vk=1000$ \kms.
{\it Thin (blue) curve:} Gaussian $N(V)$ with $\sigma=200$ \kms.
}
\label{fig:spectra}
\end{figure}

Figure~\ref{fig:nofv} shows $N(V)$ for bound clouds
with $\gamma=1$ and $2$,
as seen from a direction perpendicular to the kick.
(This is the a priori most likely direction for observing a prolate
object. Since the \rgc\ is nearly spherical within a few $\rk$,
the results cited below depend weakly on viewing angle.)
Since more than 1/2 of the stars lie at $r>2\rk$
and are moving with $v<\vk$,
the central core of the distribution has an effective width
that is much  smaller than $\vk$;
most of the information about the high velocity stars near 
the SMBH is contained in the extended wings
\citep[e.g.][]{Marel:94}.

Velocity dispersions of stellar systems are typically measured
by comparing an observed, absorption line spectrum with
template spectra that have been broadened with Gaussian $N(V)$'s; 
the comparison is either made directly in intensity-wavelength space
\citep[e.g.][]{Morton:73} or via cross-correlation \citep[e.g.][]{Simkin:74}.
For example, internal velocities of UCDs (ultra-compact dwarf galaxies) 
in the Virgo and Fornax clusters have been determined in both ways
%\citep[e.g.][]{Hilker:07 Mieske:08}.
(e.g. Hilker et al. 2007; Mieske et al. 2008).
Figure~\ref{fig:spectra} shows the results of broadening the
spectrum of a K0 star in the CaII triplet region 
($8400$\AA $\le\lambda\le 8800$ \AA),
with two broadening functions: $N(V)$ from the top panel of
Figure~\ref{fig:nofv}, scaled to $\vk=1000$ \kms,
and a Gaussian $N(V)$ with $\sigma=200$ \kms.
The two broadening functions produce similar changes
in the template spectrum; the $N(V)$ from the bound cloud generates
more `peaked' absorption lines, but this difference would be difficult
to see absent very high quality data.

We computed the best-fit, Gaussian $\sigma$ corresponding to the
various broadening functions in Figure~\ref{fig:nofv} as a function
of $\vk$.
The stellar template of Figure~\ref{fig:spectra} was convolved 
with Gaussian $N(V)$'s having $\sigma$ in the range $2$ to $2000$ 
\kms and a step size of $1$ \kms.
Each of the Gaussian-convolved templates was then compared with the
simulated \rgc~ spectrum, and
the ``observed'' velocity dispersion $\sigma_{\rm obs}$
was defined as the $\sigma$ for which the Gaussian-convolved 
template was closest, in a least-squares sense, to the \rgc~ spectrum.
No noise was added to either the \rgc~ or comparison spectra.

\begin{figure}
\includegraphics[clip,angle=-90.,width=0.45\textwidth]{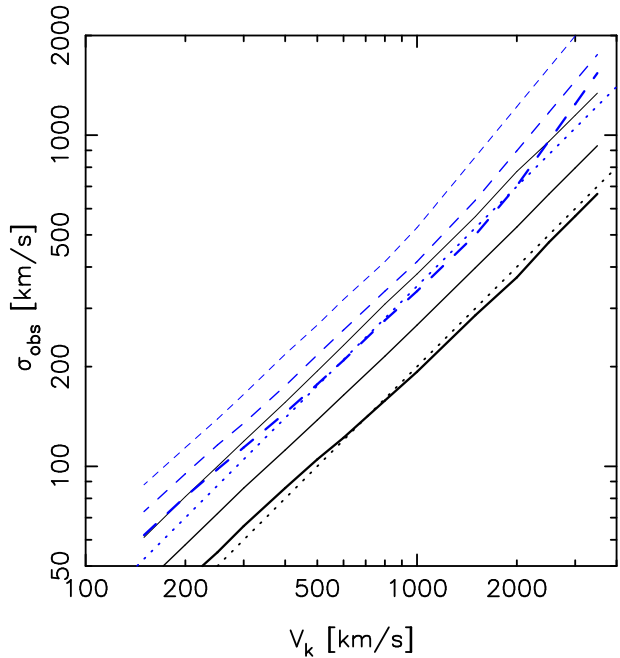}
\caption{Velocity dispersions $\sigma_{\rm obs}$ that would 
be inferred from broadened absorption-line spectra of \rgc's.
{\it Solid (black) lines:} $\gamma=1$;
{\it dashed (blue) lines:} $\gamma=2$, where $\gamma$ is the power-law
index of the stellar density profile before the kick.
Thick curves correspond to all bound stars;
thinner curves correspond to an observing aperture that includes 
only bound stars within a projected distance $10\rk$ and $3\rk$
from the SMBH (as viewed from a direction perpendicular to the kick).
Dotted lines show $\sigma_{\rm obs}=0.2\vk$ and $\sigma_{\rm obs}=0.35\vk$.
}
\label{fig:compare}
\end{figure}

\begin{figure}
\includegraphics[clip,width=0.45\textwidth]{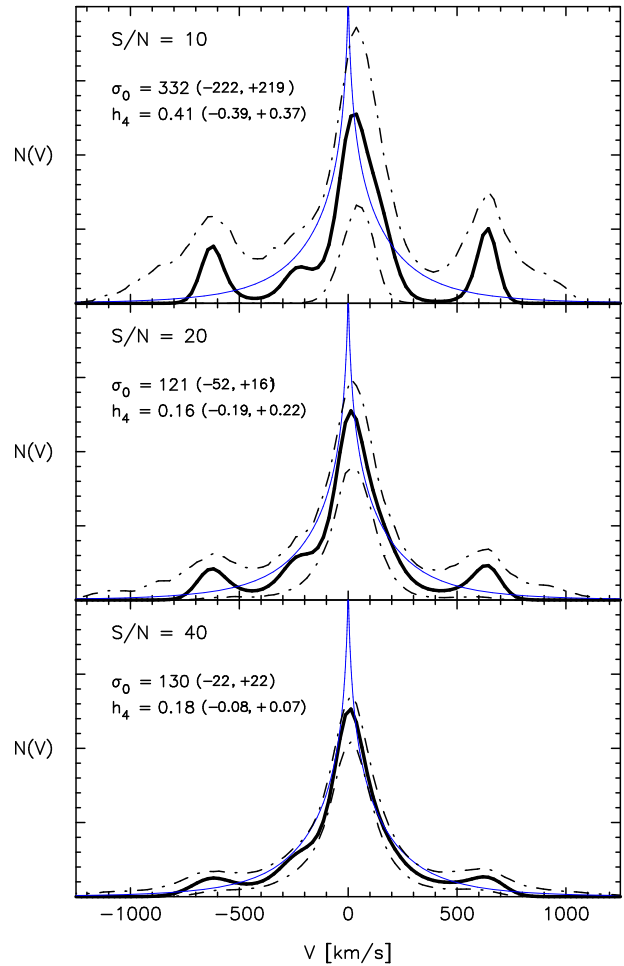}
\caption{Recovery of \rgc~broadening functions from simulated
absorption line spectral data with various amounts of added noise.
Blue lines are the input $N(V)$ (from Figure~\ref{fig:nofv}, 
with $\gamma=1$ and $\vk=10^3$ \kms).
Solid lines are the recovered $N(V)$s and dash-dotted lines are
90\% confidence bands.
$\sigma_0$ and $h_4$ are coefficients of the Gauss-Hermite
fit to the recovered $N(V)$; 90\% confidence intervals on the
parameters are given.
\label{fig:mpl}}
\end{figure}

Figure~\ref{fig:compare} shows the results for $\gamma=(1,2)$,
$150$ \kms $\le \vk \le 4000$ \kms, 
and for (circular) apertures of various sizes.
When the entire \rgc~ is included in the slit,
$\sigma_{\rm obs}\approx 0.15\vk$ ($\gamma=0.5$),
$\approx 0.20\vk$ ($\gamma=1$), and $\approx 0.35\vk$ ($\gamma=2$).
These values are well fit by the ad hoc relation
\beq
\ln F_3 = -2.17 + 0.56\gamma, \ \ \ 0.5\lap\gamma\lap 2.
\label{eq:sofgamma}
\eeq
As the aperture is narrowed, $\sigma_{\rm obs}$ increases to values
closer to $\vk$, although as argued above, realistic slits would
be expected to include essentially the entire \rgc\ and we will
assume this in what follows.

We note that some ultra-compact dwarf galaxies (UCDs) 
have $\sigma_{\rm obs}$ as large as
$40-50$ \kms and that the implied masses are difficult to reconcile
with simple stellar population models, which has led to suggestions
that the UCDs are dark-matter dominated
(Hilker et al. 2007; Mieske et al. 2008).
Alternatively, some UCD's might be bound by a central black hole;
for instance, an observed $\sigma$ of $50$ \kms is consistent with 
an \rgc~ produced via a kick of  $\sim 250$ \kms ($\gamma=1$).
Detection of the high-velocity wings in $N(V)$ (Fig.~\ref{fig:nofv})
could distinguish between these two possibilities.

While spectral deconvolution schemes exist that can
do  this \citep[e.g.][]{Saha:94,Merritt:97}, 
they require high signal-to-noise ratio data.
Precisely how high is suggested by Figure~\ref{fig:mpl},
which shows the results of simulated recovery of \rgc~broadening 
functions from absorption line spectra.
The spectrum of Figure~\ref{fig:spectra} was convolved with 
the $\gamma=1$ $N(V)$ plotted in Figure~\ref{fig:nofv}, 
with $\vk=10^3$ \kms.
Noise was then added to the broadened spectrum 
(as indicated in the figures by the signal-to-noise ratio S/N)
and the broadening function was recovered via a non-parametric algorithm
\citep{Merritt:97}; confidence bands were constructed via the
bootstrap.
Figure~\ref{fig:mpl} suggests that S/N$\approx 40$ permits
a reasonably compelling determination of a non-Gaussian $N(V)$.
This conclusion is reinforced by the inferred values of 
the Gauss-Hermite (GH) moments $\sigma_0$ and $h_4$; the former
measures the width of the Gaussian term in the GH 
expansion of $N(V)$ while
$h_4$ measures symmetric deviations from a Gaussian.
For S/N$=40$, the recovered $h_4=0.18\pm 0.1$ (90\%),
significantly different from zero.
(We note that the velocity dispersion corresponding to the
GH expansion is $\sigma_{\rm c}=(1+\sqrt{6}h_4)\sigma_0$
which is close to $\sigma_{\rm obs}$ as defined above.)
In \S 7 we discuss the feasibility of obtaining \rgc~spectra
with such high S/N.

Combining equations~(\ref{eq:f2}) and (\ref{eq:defs}),
the (stellar) mass-velocity dispersion ($\mb-\sigma_{\rm obs}$) 
relation for \rgc's becomes
\beq
{\mb\over\msun} \approx  F_1(\gamma)\times F_3(\gamma)^{2(3-\gamma)} 
\left({\mh\over\msun}\right) 
\left({G\mh\over \rh}\right)^{3-\gamma}
\sigma_{\rm obs}^{2(\gamma-3)}.
\label{eq:msigma}
\eeq

It is tempting (though only order-of-magnitude correct) to write 
$\rh\approx r_{\rm infl}\approx G\mh/\sigma^2$, which allows 
equation~(\ref{eq:msigma}) to be written
\beq
{\mb\over\mh} \approx  F_1\times F_3^{2(3-\gamma)} 
\left({\sigma_{\rm obs}\over\sigma}\right)^{2(\gamma-3)}.
\label{eq:fake}
\eeq
The dimensionless coefficient in these two expressions is equal to
$(3\times 10^{-3},1.2\times 10^{-2},0.4)$
for $\gamma=(0.5,1,2)$.

\bigskip\bigskip

\section{The Pre-Kick Stellar Density}
While the linear extent of a \rgc~ is determined entirely by $\mh$ and $\vk$
(eq.~\ref{eq:rk}),
its luminosity and (stellar) mass depend also on the density of stars
around the SMBH (i.e.\ around the massive binary) just prior to the kick.
In this section we discuss likely values for the parameters
that determine the pre-kick density of stars near the SMBH and
the implications for  the mass that remains bound after the kick.
In a following section we will relate mass to luminosity and color.

Two inspiralling SMBHs first form a bound pair when
their separation falls to $\sim r_{\rm infl}\equiv G\mh/\sigma$, 
the influence radius of the larger hole.
This distance is a few parsecs in a galaxy like the Milky Way.
The separation between the two SMBHs then drops very rapidly 
(on a nuclear crossing time scale) to a fraction $\sim 0.1 M_2/M_1$ 
of $r_{\rm rinfl}$  as the 
binary kicks out stars on intersecting orbits via the gravitational
slingshot \citep{Merritt:06}.
Because a massive binary tends to lower the density of stars or gas
around it, the two SMBHs may stall at this separation,
never coming close enough together
($\lap 10^{-3}$ pc) that gravitational wave emission can
bring them to full coalescence.
This is the ``final parsec problem.''

Of course, in order for a kick to occur, the two SMBHs {\it must}
coalesce, and in a time shorter than $\sim 10$ Gyr.
Roughly speaking, this requires that the density of stars or
gas near the binary remain high until shortly before coalescence.
This implies, in turn, a relatively large mass in stars that 
can remain bound to the SMBH after the kick, hence a relatively
large luminosity for the \rgc~ that results.

Converting these vague statements into quantitative estimates
of the stellar density just before the kick
requires a detailed model for the joint evolution
of stars and gas around the shrinking binary.
A number of such models have been discussed
\citep[see][for a review]{GM:08}.
Here we focus on the two that are perhaps best understood:
\begin{itemize}
\item {\it Collisional loss-cone repopulation.} If the
two-body relaxation time $t_R$ in the pre-kick nucleus is 
sufficiently short, gravitational scattering between stars can
continually repopulate orbits that were depleted by the
massive binary, allowing it to shrink on a timescale
of $\sim t_R$ \citep{Yu:02}.
This process can be accelerated if the nucleus contains 
perturbers that are significantly more massive than stars,
e.g. giant molecular clouds \citep{Perets:08}.
Repopulation of depleted orbits guarantees that the density
of stars near the binary will remain relatively high as 
the binary shrinks.
\item {\it Collisionless loss-cone repopulation.} In non-axisymmetric
(barred, triaxial or amorphous) galaxies, some orbits are 
``centrophilic,'' passing near the galaxy center each crossing
time.
This can imply feeding rates to a central binary 
as large as $\dot M\sim G^{-1}\sigma^3$ even in the absence
of collisional loss-cone repopulation \citep{MP:04}.
Because the total mass on centrophilic orbits can be $\gg\mh$, 
interaction of the binary with a mass $\sim \mh$ in stars need
not imply a significiant decrease in the local density of stars,
again implying a large pre-kick density near the binary.
\end{itemize}
\noindent
We now discuss these two pathways in more detail and their
implications for the pre-kick stellar density near the SMBH.

\subsection{Collisional loss-cone repopulation}

At the end of the rapid evolutionary phase described above,
the binary forms a bound pair with semi-major axis
\beq
a\approx a_{\rm h} \equiv {q\over (1+q)^2} {r_{\rm infl}\over 4}\, ,
\label{eq:ah}
\eeq
with $q\equiv M_2/M_1\le 1$ the binary mass ratio
\citep[e.g.][]{Merritt:06}.
Stars on ``loss cone'' orbits that intersect the binary 
have already been removed via the gravitational slingshot by this time, 
and continued evolution of the binary is determined by the rate at
which these orbits are repopulated -- in this model, via
gravitational scattering.
Scattering onto loss-cone orbits around a central 
mass $\mh=M_1+M_2$ occurs predominantly from stars on
eccentric orbits with semi-major axes $\sim r_{\rm infl}$,
and the relevant relaxation time is therefore
$\sim t_R(r_{\rm infl})$.
Relaxation times at $r=r_{\rm infl}$ in real galaxies 
are found to be well correlated with spheroid luminosities
\cite[e.g. Figure~4 of][]{MHB:07},
dropping below $10$ Gyr only in low-luminosity spheroids -- 
roughly speaking, fainter than the bulge of the Milky Way.
Such spheroids have velocity dispersions $\lap 150$ \kms
and contain SMBHs with masses $\lap 10^7\msun$.
Binary SMBHs in more luminous galaxies might still evolve
to coalescence via this mechanism, but only if they contain significant 
populations of perturbers
more massive than $\sim \msun$, e.g. giant molecular clouds
or intermediate mass black holes;
Perets \& Alexander (2008) have argued that this might generically be the
case in the remnants of gas-rich galaxy mergers though this model is
unlikely to work in gas-poor, old systems like giant elliptical galaxies.

Denoting the semi-major axis of the massive binary by $a(t)$, one finds 
\citep{MMS:07}
\beq
{1\over t_{\rm R}(r_{\rm infl})}\left|{a\over\dot a}\right| 
\approx A\ln\left({a_{\rm h}\over a}\right) + B
\eeq
for $a_{\rm eq}\lap a(t)\lap a_{\rm h}$, 
where $a_{\rm eq}\approx 10^{-3}r_{\rm infl}$ 
is the separation at which energy losses due to gravitational wave
emission begin to dominate losses due to interaction with stars;
($A,B$) $\approx$ ($0.016,0.08$) with only a weak dependence on binary 
mass ratio.
The elapsed time between $a=a_{\rm h}$ and $a=a_{\rm eq}$ is 
of order $t_R(r_{\rm infl})$.

\begin{figure}
\includegraphics[clip,angle=-90.,width=0.45\textwidth]{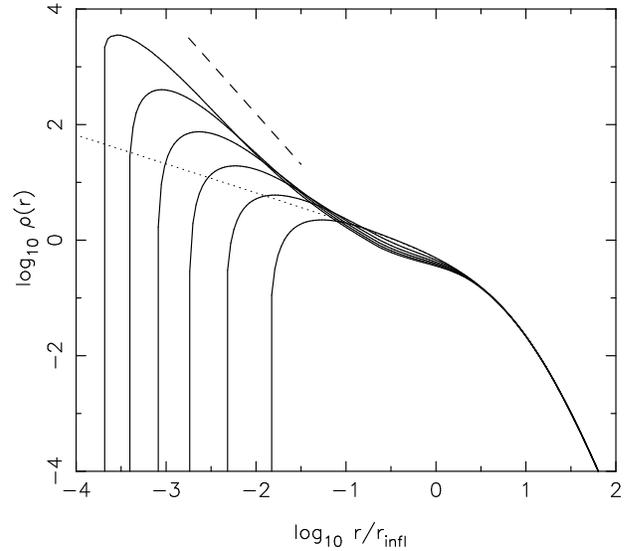}
\caption{Evolving stellar density around a binary SMBH
of mass $M_1+M_2=10^6\msun$ in a spherical galaxy containing
$10^9$ Solar-mass stars, in the ``collisional loss cone repopulation'' 
regime \citep{MMS:07}.
Solid lines show $\rho(r)$ at five different times, 
between $a(t)\approx a_{\rm h}$ and $a(t)\approx a_{\rm eq}$.
The density falls to zero at $r\approx a(t)$ and
smaller values of $a$ correspond to later times;
total elapsed time is $\sim 0.5 t_R(r_{\rm infl})$
where $r_{\rm infl}$ is the gravitational influence
radius of the massive binary.
Dotted line shows the initial (pre-binary) galaxy density
and dashed line has the Bahcall-Wolf (1976) slope,
$\rho\propto r^{-7/4}$.
}
\label{fig:coll}
\end{figure}

Figure~\ref{fig:coll} shows the evolution of the stellar density
around a massive binary as it shrinks from $a\approx a_{\rm h}$ to
$a\approx a_{\rm eq}$;
the evolution was computed using
the Fokker-Planck formalism described in Merritt et al. (2007b).
The same gravitational encounters that scatter stars into the binary
also drive the distribution of stellar energies toward the Bahcall-Wolf (1976)
``zero-flux'' form, $\rho\sim r^{-7/4}$, and on the same time scale,
$\sim t_R(r_{\rm infl})$;
as a result, a high density of stars is 
maintained at radii $a(t)\lap r \lap 0.2 r_{\rm infl}$.
In effect, the inner edge of the cusp follows the binary as the 
binary shrinks.

Once $a(t)$ drops below $\sim a_{\rm eq}$, the binary 
``breaks free'' of the stars and evolves rapidly toward coalescence,
leaving behind a phase-space gap corresponding to orbits with pericenters
$\lap a_{\rm eq}$.
(In a similar way, evolution of a binary SMBH in response to
gravitational waves and gas-dynamical 
torques leaves behind a gap in the gaseous accretion
disk; Milosavljevic \& Phinney 2005.) 
Gravitational scattering will only partially refill this gap  
in the time between $a=a_{\rm eq}$ and coalescence \citep{MW:05}.
Figure~\ref{fig:coll} suggests that $a_{\rm eq}\ll (a_{\rm h}, r_{\rm infl})$.
Merritt et al. (2007b) estimated,
based on the same Fokker-Planck model used to construct Figure~\ref{fig:coll},
that
\beq
{a_{\rm eq}\over r_{\rm infl}} \approx (0.20,0.67,2.3,7.8)\times 10^{-3}
\eeq
for equal-mass binaries with total mass 
$M_1+M_2=(10^5,10^6,10^7,10^8)M_\odot$;
the numbers in parentheses decrease by $\sim 25\%$ for binaries
with $M_2/M_1=0.1$.

\begin{figure}
\includegraphics[clip,angle=-90.,width=0.47\textwidth]{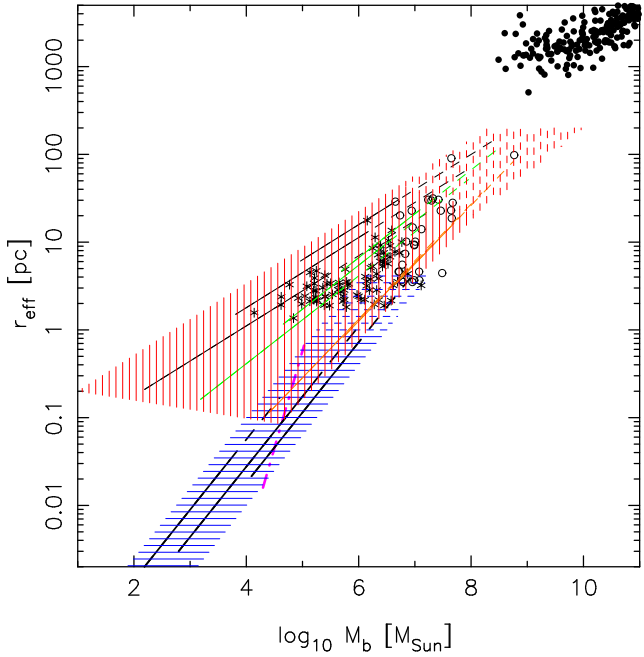}
\caption{Effective radius vs.\ bound stellar mass for \rgcs.
Thick solid lines (blue hatched area)
are based on the ``collisional'' 
loss cone repopulation model
and assume a galaxy central velocity dispersion of 
$\sigma=(50,100,150)$ \kms (from left to right).
Thin solid lines (red hatched area)
are based on the ``collisionless'' loss cone
repopulation model; the three lines in each set assume 
a galaxy central velocity dispersion of $\sigma=(200,300,400)$ \kms
(right to left) and the three sets of lines are for 
$\gamma=0.5$ (black) , 1.0 (green) and 1.5 (orange).
For both models,
solid lines extend to a maximum $r_{\rm eff}$ based on the
assumption that $\vk\ge 4.5\sigma$ (escape from the galaxy) 
while dashed lines correspond to the weaker condition
$\vk\ge 2\sigma$ (escape from the galaxy core).
\rgcs\ to the left of the dash-dotted (magenta) line are expected
to expand appreciably over their lifetime. 
Data points are from Forbes et al. (2008).
Filled circles: E galaxies.
Open circles: Ultra-compact Dwarfs (UCDs) and Dwarf-Globular
Transition Objects (DGTOs).
Stars: globular clusters.
}
\label{fig:RM}
\end{figure}

\begin{figure}
\includegraphics[clip,angle=-90.,width=0.47\textwidth]{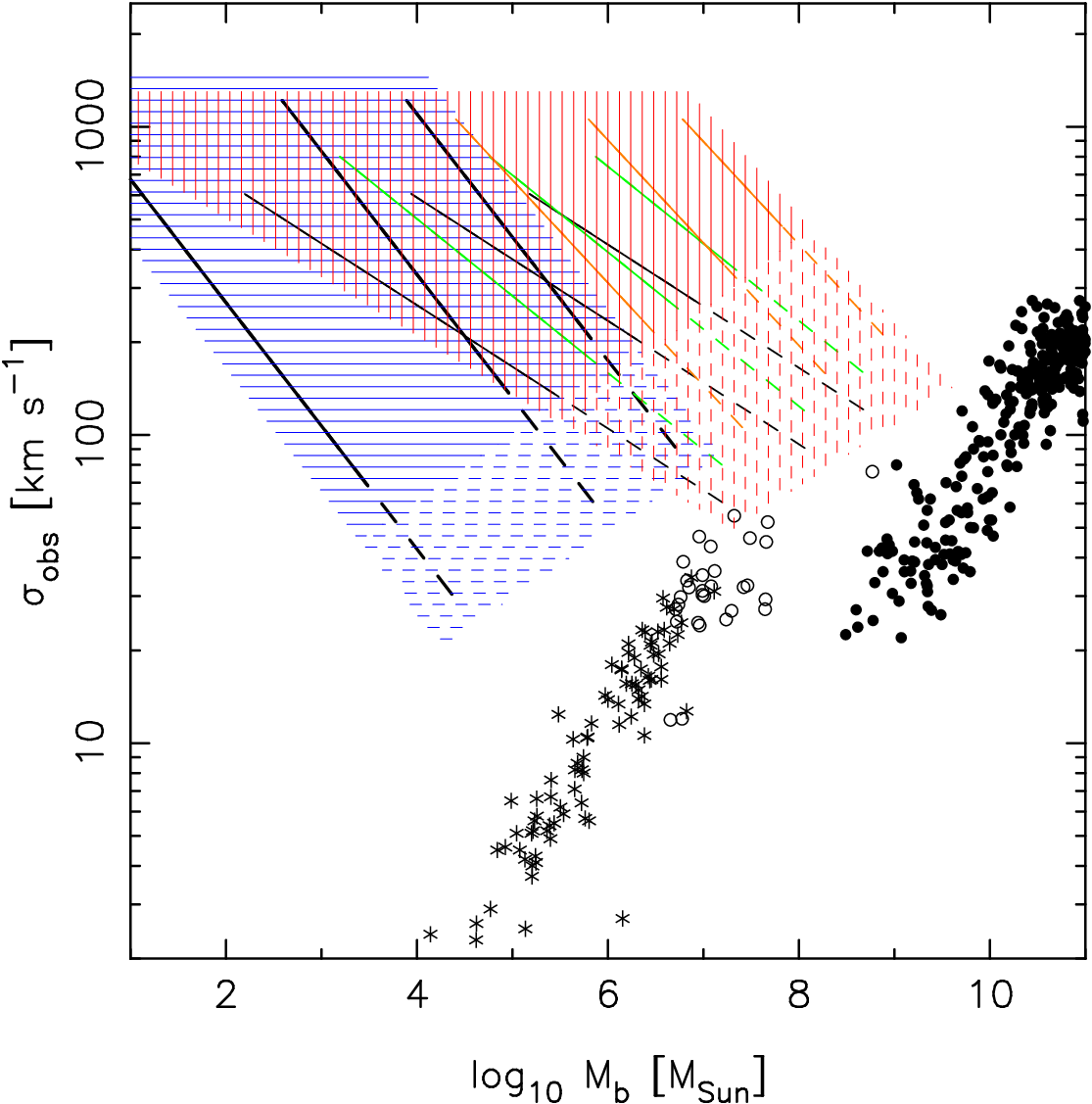}
\caption{Observed velocity dispersion vs.\ bound stellar mass for \rgcs.
Thick solid lines (blue hatched area) are based on the 
``collisional'' loss cone repopulation model
and assume a galaxy central velocity dispersion of 
$\sigma=(50,100,150)$ \kms (from left to right).
Thin solid lines (red hatched area) are based on the 
``collisionless'' loss cone
repopulation model; the three lines in each set assume 
a galaxy central velocity dispersion of $\sigma=(200,300,400)$ \kms
(right to left) and the three sets of lines are for 
$\gamma=0.5$ (black) , 1.0 (green) and 1.5 (orange).
For both models,
solid lines extend to a minimum $\sigma_{\rm obs}$ based on the
assumption that $\vk\ge 4.5\sigma$ (escape from the galaxy) 
while dashed lines correspond to the weaker condition
$\vk\ge 2\sigma$ (escape from the galaxy core).
Other symbols are as in Fig.~\ref{fig:RM}.
}
\label{fig:SM}
\end{figure}

Following the kick, the density profile of Figure~\ref{fig:coll} 
will be truncated beyond $r\approx\rk$.
The inner cutoff at $r\approx a_{\rm eq}$ satisfies
\beq
{a_{\rm eq}\over \rk} \approx 10^{-3} \left({\vk\over\sigma}\right)^2.
\eeq
The requirement that $a_{\rm eq}<\rk$ 
-- i.e. that at least {\it some} stars remain bound after the kick
-- then becomes $\vk\lap 30 \sigma$,
which is never violated by reasonable ($\vk,\sigma$) values.
However, the inner cutoff exceeds $0.1\rk$ for
$\vk\gap 10\sigma$, a condition that would be fulfilled
for $\sigma=100$ \kms and $\vk\gap 10^3$ \kms.
In what follows we ignore the inner cutoff and assume that
the Bahcall-Wolf cusp extends to $r=0$.

The pre-kick density can therefore be approximated as
\begin{equation}\label{eq:rho}
\rho(r) = \left\{ \begin{array}{lcc}
  \rho_0\left({r\over r_0}\right)^{-7/4} & : & 0\lap r \lap r_0 \\
  \rho_0 & : & r_0\lap r \lap r_{\rm infl} \end{array} \right. ,
\end{equation}
where $r_0\approx 0.2 r_{\rm infl}$ and $\rho_0=\rho(r_0)$ 
is the density of the galaxy core.
Using equations (\ref{eq:f1}) and (\ref{eq:rho}),
the mass remaining bound to the coalesced SMBH after a kick is then
\begin{subequations}
\begin{eqnarray}
{\mb\over\mh} &=& F_1(\gamma)
\left({G\mh\over \rh\vk^2}\right)^{3-\gamma} \\
&\approx& 1.3 \left({G\mh\over r_{\rm infl}\vk^2}\right)^{5/4} 
\left({\rho_0r_{\rm infl}^3\over\mh}\right) \\
&\approx& \left({\sigma\over\vk}\right)^{5/2} 
\left({M_{\rm core}\over\mh}\right) \, ,
\label{eq:m3}
\end{eqnarray}
\end{subequations}
where $M_{\rm core}\equiv\rho_0r_{\rm infl}^3$;
the last expression assumes $\rk<0.2r_{\rm infl}$, 
i.e. $\vk\gap 2\sigma$, which is always satisfied for a \rgc\
that escapes the galaxy core.

\begin{figure}
\includegraphics[clip,angle=-90.,width=0.45\textwidth]{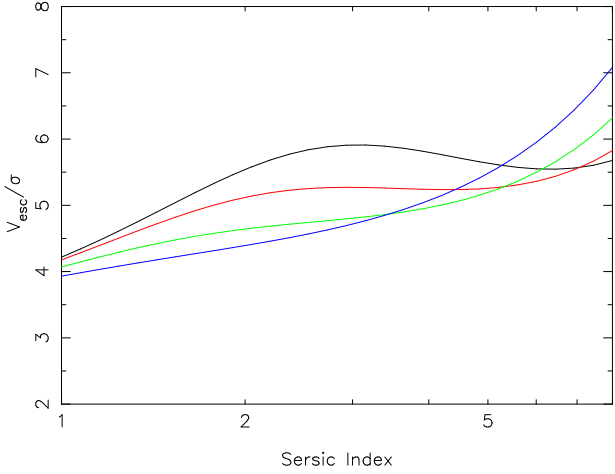}
\caption{Escape velocity from the center of a Sersic-law galaxy as a function
of Sersic index $n$, in units of the central, projected, 1d velocity
dispersion as measured through a circular aperture.
Constant mass-to-light ratio was assumed and the effect of the SMBH
on the potential or on the motions of stars was ignored in 
computing $V_{\rm esc}$ and $\sigma$.
The four curves (black, red, green, blue) correspond to aperture radii 
of (0.01,0.03,0.1,0.3) in units of the half-light radius of the galaxy.
}
\label{fig:escape}
\end{figure}

To the extent that the galaxy core was itself created by the massive
binary during its rapid phase of evolution, then
$M_{\rm core}/\mh$ is of order unity \citep{Merritt:06}.
(Following the kick, the core will expand still more; 
Gualandris \& Merritt 2008.)
Making this assumption yields
\begin{subequations}
\begin{eqnarray}
&&{\mb\over\mh} \approx 
\left({\sigma\over\vk}\right)^{5/2} \\
&&\approx 2\times 10^{-2} 
\left({\sigma\over 200 \mbox{ km s}^{-1}}\right)^{5/2} 
\left({\vk\over 10^3 \mbox{ km s}^{-1}}\right)^{-5/2}. 
\end{eqnarray}
\label{eq:m4}
\end{subequations}
In the case of ejection from a stellar spheroid like that
of the Milky Way
($\sigma\approx 100$ \kms, $\mh\approx 4\times 10^6\msun$),
we have
\beq
\mb \approx 10^4 \msun \left({\vk\over 10^3 {\rm km\ s}^{-1}}\right)^{-5/2},
\eeq
i.e. 
\begin{subequations}
\begin{eqnarray}
3\times 10^2 \lap &\mb/\msun& \lap 5\times 10^4, \\
4000 \ge &\vk/({\rm km\ s}^{-1})& \ge 500.
\end{eqnarray}
\end{subequations}

Combining equations~(\ref{eq:rk}), (\ref{eq:f2}) and (\ref{eq:m4})  
gives the mass-radius relation in the collisional regime: 
\begin{subequations}
\begin{eqnarray}
\mb  
&&\approx 0.4 G^{-5/4}\mh^{-1/4}\sigma^{5/2}r_{\rm eff}^{5/4} \\
%&&\approx 0.4 G^{-1}r_{\rm infl}^{-1/4}\sigma^2r_{\rm eff}^{5/4} \\
&&\approx 4\times 10^4 \msun \left({\mh\over 10^7\msun}\right)^{-1/4}
\left({\sigma\over 100\ {\rm km\ s}^{-1}}\right)^{5/2}
\left({r_{\rm eff}\over 0.1\ {\rm pc}}\right)^{5/4}\\
&&\approx 5\times 10^4 \msun 
\left({\sigma\over 100\ {\rm km\ s}^{-1}}\right)^{1.29}
\left({r_{\rm eff}\over 0.1\ {\rm pc}}\right)^{5/4}\, ,
\end{eqnarray}
\label{eq:rmcol}
\end{subequations}
where the final expression assumes the $\mh-\sigma$ relation in equation
(\ref{eq:MS}).

The mass-velocity dispersion ($\mb-\sigma_{\rm obs}$) relation 
for \rgcs\ follows from 
equation~(\ref{eq:m4}) with $\sigma_{\rm obs} = F_3(\gamma)\vk\approx 0.30\vk$:
\begin{subequations}\label{eq:Mb_collisional}
\begin{eqnarray}
\mb &\approx& 0.05 \mh \sigma^{5/2} \sigma_{\rm obs}^{-5/2} \\
&\approx& 2\times 10^5\msun 
\left({\sigma\over 100\ {\rm km\ s}^{-1}}\right)^{7.4}
\left({\sigma_{\rm obs}\over 100\ {\rm km\ s}^{-1}}\right)^{-5/2}
\end{eqnarray}
\label{eq:smcol}
\end{subequations}
where the $\mh-\sigma$ relation has again been used.

Figures~\ref{fig:RM} and~\ref{fig:SM} plot the relations
(\ref{eq:rmcol}), (\ref{eq:smcol}) for $\sigma=(50,100,150)$ \kms.
Plotted for comparison are samples of globular clusters and dwarf
galaxies from the compilation of Forbes et al. (2008).

In these plots, the minimum $r_{\rm eff}$ is presumed to be
that associated with a kick of $\sim 4000$ \kms.
This condition (combined with the $\mh-\sigma$ relation) gives
\beq
r_{\rm eff} \gap 3.0\times 10^{-3}\, {\rm pc} \left({\sigma\over 100\ {\rm km\ s}^{-1}}\right)^{4.86}.
\eeq
The maximum $r_{\rm eff}$ is associated with the smallest 
$\vk$ that is of physical interest. 
We express this value of $\vk$ as $N\sigma$ which allows us to write
\beq\label{eq:Nsigma}
r_{\rm eff} \lap 4.9\, {\rm pc}\, N^{-2} 
\left({\sigma\over 100\ {\rm km\ s}^{-1}}\right)^{2.86}.
\eeq
Kicks large enough to eject a SMBH completely from its galaxy have
$N\approx 5$ (Figure~\ref{fig:escape}).
Kicks just large enough to remove a SMBH from the galaxy
core have $N\approx 2$ \citep{GM:08}.
Figures~\ref{fig:RM} and~\ref{fig:SM} show the limits on $r_{\rm eff}$
and $\sigma_{\rm obs}$
corresponding to $N=5$ and $N=2$, the latter via dashed lines.

According to Figure~\ref{fig:RM}, \rgcs\ in this ``collisional''
regime can have effective radii as big as $\sim 1$ pc
when $\vk\gap V_{\rm esc}$; on the $\mb-r_{\rm eff}$ plane their 
distribution barely overlaps with globular clusters, 
and extends to much lower sizes and (stellar) masses.
However their velocity dispersions (Fig.~\ref{fig:SM})
would always substantially exceed those of either globular
clusters or compact galaxies of comparable (stellar) mass.

\subsection{Collisionless loss-cone repopulation}

By ``collisionless'' we mean that the nuclear
relaxation time is so long that gravitational scattering
can not refill the loss cone of a massive binary at a fast
enough rate to significantly affect the binary's evolution
after the hard-binary regime (eq.~\ref{eq:ah}) has been reached.
The relevant radius at which to evaluate the relaxation time 
is $\sim \rinfl$, the influence radius of the binary
(or of the single black hole that subsequently forms).
The relaxation time at $\rinfl$ in elliptical galaxies is 
found to correlate tightly with $\sigma$ or $\mh$
\citep{MMS:07}:
\begin{equation}
t_R(\rinfl) \approx 8.0\times 10^9\ {\rm yr}
\left({\mh\over 10^6 M_\odot}\right)^{1.54} ,
\label{eq:trsigma}
\end{equation}
where Solar-mass stars have been assumed.
A mass of order $\mh$ is scattered into the central sink
in a time $t_R(\rinfl)$, and this is also roughly the mass
that must interact with the binary in order for it
to shrink by a factor of order unity.
Even allowing for variance in the phenomenological relations 
(\ref{eq:trsigma}),
it follows that collisional loss cone refilling is unlikely 
to significantly affect the evolution of a binary SMBH 
in galaxies with $\sigma\gap 200$ \kms or $\mh\gap 10^8M_\odot$.

An alternative pathway exists for stars in these galaxies to
interact with a central binary.
If the large-scale galaxy  potential is non-axisymmetric,
a certain fraction of the stellar orbits will have filled centers
-- these are the (non-resonant) box or centrophilic orbits,
which are typically chaotic as well due to the presence
of the central point mass \citep{MV:99}.
Stars on centrophilic orbits pass near the central object once per
crossing time; the number of near-center passages that come
within a distance $d$ of the central object, per unit of time,
is found to scale roughly linearly with $d$ \citep{GB:85,MP:04},
allowing the rate of supply of stars to a central object
to be computed simply given the population of centrophilic orbits.
While the latter is not well known for individual galaxies,
stable, self-consistent triaxial galaxy models with central
black holes can be constructed with chaotic orbit fractions as
large as $\sim 70\%$ \citep{PM:04}.
Placing even a few percent of a galaxy's mass on centrophilic
orbits is sufficient to bring two SMBHs to coalescence
in 10 Gyr  \citep{MP:04}.
Furthermore, the effect of the binary on the density of stars
in the galaxy core is likely to be small, since the mass associated
with centrophilic orbits is  $\gg\mh$ and stars on
these orbits spend most of their time far from the center.

Here, we make the simple assumption that {\it the observed core structure
of bright elliptical galaxies is similar to what would result from
the decay and coalescence of a binary SMBH} in the collisionless
loss-cone repopulation model.
In other words, we assume that the binary SMBHs that were
once present in these galaxies {\it did} coalesce, and
the cores that we now see are relics of  the binary evolution
that preceded that coalescence.
By making these assumptions, we are probably underestimating
the density around a SMBH at the time of a kick,
since some observed cores will have been enlarged by the kick itself
\citep{GM:08}.
Also, core sizes in local (spatially resolved) galaxies 
are likely to reflect a series of past merger events
\citep{Merritt:06};
SMBHs that recoiled during a previous generation of mergers would
probably have carried a higher density of stars than implied by
the current central densities of galaxies.

Above we characterized the pre-kick mass density as 
$\rho\propto r^{-\gamma}$, with $\rh$ the radius at which
the enclosed stellar mass equals twice $\mh$.
We computed $\gamma$ and $\rh$ for a subset of early-type 
galaxies in the  ACS Virgo sample \citep{ACSI} for which
$\sigma$ was known; for some of these galaxies the SMBH
mass has been measured dynamically while $\mh$ in the
remaining galaxies was computed from equation~(\ref{eq:MS}).
Each galaxy was modelled with a PSF-convolved, 
core-Sersic luminosity profile \citep{Graham:03}, 
which assumes a power law relation between luminosity
density and projected radius inside a break radius $R_{\rm b}$.
The core-Sersic fits were numerically deprojected, and converted
from a luminosity to a mass density as in Ferrarese et al. (2006).
The radius $\rh$ was then computed from
\beq
\rh = \left( {3-\gamma\over\pi}
    {\mh\over\rho_0r_0^\gamma}\right)^{1/(3-\gamma)} \, ,
\eeq
with $\gamma$ the central power-law index of the deprojected
density; $r_0$ is a fiducial radius smaller than $R_{\rm b}$ 
which we chose to be $1$ pc and $\rho_0$ is the mass density at $r=r_0$.

\begin{figure}
\includegraphics[clip,width=0.45\textwidth]{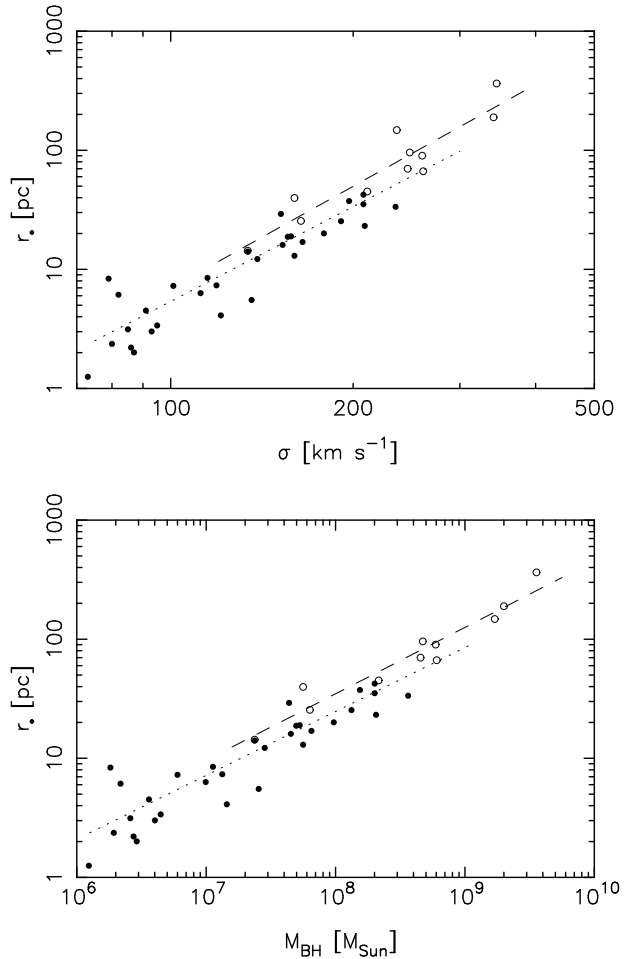}
\caption{Relation between $\rh$, the radius containing a mass
in stars equal to twice $\mh$, and central velocity
dispersion ({\it top}) or black  hole mass ({\it bottom}), for galaxies in 
the ACS Virgo sample.
Open circles are ``core'' galaxies; dashed lines are fit
to just these points while dotted lines are fit to the entire
sample.
}
\label{fig:rbh}
\end{figure}

Figure~\ref{fig:rbh} shows the relation between $\rh$ and
$\sigma$ and between $\rh$ and  $\mh$.
``Core'' galaxies (those with projected profiles flatter than
$\Sigma \propto R^{-0.5}$ near the center) are plotted with open circles
and ``power-law'' galaxies ($\Sigma$ steeper than $R^{-0.5}$) 
galaxies as filled circles.
While this distinction is somewhat arbitrary, Figure~\ref{fig:rbh}
confirms that the ``core'' galaxies have larger $\rh$ at given
$\sigma$ or $\mh$ than the ``power-law'' galaxies, consistent with
the idea that the central densities of ``core'' galaxies have
been most strongly affected by mergers.
The best-fit relations defined by the core galaxies alone are
\begin{subequations}
\begin{eqnarray}
\log_{10} (\rh/\rm pc) &=& -4.84 + 2.84 \log_{10} (\sigma/{\rm km\ s}^{-1}) \\
&=& -2.92 + 0.56 \log_{10} (\mh/M_\odot)
\end{eqnarray}
\end{subequations}
i.e.
\begin{subequations}
\begin{eqnarray}
\rh &\approx& 50\ {\rm pc} \left({\sigma\over 200\ {\rm km\ s}^{-1}}\right)^{2.8} \\
&\approx& 35\ {\rm pc} \left({\mh\over 10^8\msun}\right)^{0.56}.
\end{eqnarray}
\label{eq:rhcl}
\end{subequations}
While these relations are fairly tight, the $\gamma$ values show
somewhat more scatter, in the range $0.5\lap\gamma\lap 1.5$,
and we leave $\gamma$ as a free parameter in what follows.

Combining the relations (\ref{eq:rhcl}) with equations~(\ref{eq:mvsr}) 
and (\ref{eq:MS})
gives a mass-radius relation for \rgcs\ in the ``collisionless'' paradigm:
\begin{subequations}
\begin{eqnarray}
{\mb\over 10^4\msun} &\approx& G_{\rm s}(\gamma) 
\left({\sigma\over 200\ {\rm km\ s}^{-1}}\right)^{2.84\gamma-3.66}
\left({r_{\rm eff}\over 0.1\ {\rm pc}}\right)^{3-\gamma},    \\
&\approx& G_{\rm m}(\gamma) 
\left({\mh\over 10^8\msun}\right)^{0.560\gamma-0.680}
\left({r_{\rm eff}\over 0.1\ {\rm pc}}\right)^{3-\gamma},
\end{eqnarray}
\label{eq:G1}
\end{subequations}
where
\begin{subequations}
\begin{eqnarray}
G_{\rm s}(\gamma) &=& 1.93\times 10^5 \gamma^{-1.75} 
\left(750.\over 2^{1/(3-\gamma)}-1\right)^{\gamma-3}, \\
G_{\rm m}(\gamma) &=& 1.16\times 10^5 \gamma^{-1.75} 
\left(525.\over 2^{1/(3-\gamma)}-1\right)^{\gamma-3}.
\end{eqnarray}
\label{eq:G2}
\end{subequations}

Similarly, combining equations~(\ref{eq:rhcl}) with equation~(\ref{eq:msigma})
gives the mass-velocity dispersion relation:
\begin{subequations}\label{eq:Mb_collisionless}
\begin{eqnarray}
{\mb\over 10^4\msun} &\approx& H_{\rm s}(\gamma) 
\left({\sigma\over 200\ {\rm km\ s}^{-1}}\right)^{10.92 - 2.02\gamma}
\left({\sigma_{\rm obs}\over 100\ {\rm km\ s}^{-1}}\right)^{2(\gamma-3)} \\
&\approx& H_{\rm m}(\gamma)
\left({\mh\over 10^8\msun}\right)^{2.32-0.44\gamma}
\left({\sigma_{\rm obs}\over 100\ {\rm km\ s}^{-1}}\right)^{2(\gamma-3)}
\end{eqnarray}
\label{eq:H1}
\end{subequations}
where
\begin{subequations}
\begin{eqnarray}
H_{\rm s}(\gamma) &=& 1.92\times 10^5 \gamma^{-1.75}[1.2F_3(\gamma)]^{2(3-\gamma)}, \\
H_{\rm m}(\gamma) &=& 1.16\times 10^5 \gamma^{-1.75}[1.1F_3(\gamma)]^{2(3-\gamma)}
\end{eqnarray}
\label{eq:H2}
\end{subequations}
and $S(\gamma)$ is given by equation~(\ref{eq:sofgamma}).

These relations are plotted in Figures~\ref{fig:RM} and~\ref{fig:SM}.
The allowed locus in the $r_{\rm eff} - \mb$ diagram (indicated by red
vertical lines) now includes
the region occupied also by globular clusters and compact E galaxies.
However, velocity dispersions remain much higher than those observed
so far in these classes of object.

\subsection{Unbound stars}

The high, pre-kick stellar densities near the binary which are
required for coalescence in the stellar-dynamical models
do not necessarily imply that these stars are 
bound to the SMBH.
For instance, in a triaxial galaxy populated by 
radially-anisotropic box orbits, 
some of the stars that are momentarily  near SMBH will be on orbits 
that make them unbound with respect to the hole, even
before  the kick.
Another example is loss-cone repopulation by ``massive perturbers'';
in this model stars are ``shot'' inward to the binary on eccentric
orbits, many with high enough velocities that they would be unbound
in the absence of the galactic potential.

When deriving the velocity distribution of stars near the SMBH,
we assumed isotropy and we neglected the effect of the galaxy's
gravitational potential on the stellar orbits.
These assumptions would be violated in some (though not all)
of  the loss-cone repopulation mechanisms that have been invoked
to solve the stalling problem.
Here we discuss briefly the  consequences of relaxing these assumptions.

We first note that in the {\it isotropic} case,
unbound stars are negligibly important.
We verified this by constructing fully self-consistent 
models of galaxies containing SMBHs and counting the unbound stars at each 
radius. We used (isotropic) Dehnen (1993)
models, which have an inner power-law density profile; the self-consistent
distribution function describing the stars was computed assuming a
central point with mass $\mh=0.002 M_{\rm gal}$. We confirmed, for
$\gamma=1$ and $\gamma=2$, that the fraction of the mass within
$\rk$ that is unbound for kick velocities exceeding the escape velocity
is never more than a few percent.

If the pre-kick velocity distribution were radially anisotropic,
more stars would be unbound with respect to the massive binary
and the post-kick bound population would be smaller than what was computed
in \S 2.
While this is certainly possible, we note that the anisotropy would have to be 
appreciable at very small radii, a tenth or hundredth of the SMBH influence 
radius, in order to substantially affect the fraction of the population
that is bound after the kick.  

For $\gamma < 0.5$, there is no isotropic distribution function that can 
reproduce the density near the SMBH (or rather, the isotropic distribution
function would be negative at highly-bound energies).
For such flat cores, one would need to assume a {\it tangentially} anisotropic 
velocity distribution. This  would  have the  effect of {\it increasing}
the mass of the post-kick
bound population, since it would increase the number of stars on low-velocity 
(nearly circular) orbits at every radius. 

These uncertainties deserve to be more completely addressed in a future
paper. For now, we prefer to encapsulate them all in the value of
$\gamma$.

\subsection{Other pathways to coalescence and/or ejection}
Recoils large enough to eject SMBHs completely from galaxies
can occur even in the absence of gravitational waves,
via Newtonian interactions involving
three or more massive objects \citep[e.g.][]{MV:90,HL:06}.
The same interactions can also hasten coalescence of a binary
SMBH by inducing changes in its orbital eccentricity.
If the infalling SMBH is less massive than either
of the components of the pre-existing binary,
$M_3<(M_1,M_2)$,
the ultimate outcome is likely to be ejection of the 
smaller hole and recoil of the binary, with the binary
eventually returning to the galaxy center.
If $M_3>M_1$ or $M_3>M_2$, there will most often
be an exchange interaction, with the lightest SMBH
ejected and the two most massive SMBHs forming a
binary; further interactions then proceed as in 
the case $M_3<(M_1,M_2)$.
Whether, and to what extent, the ejected SMBH constitutes
a HCSS depends on whether it carries a bound population
and can retain it during interaction with the other SMBHs.
These questions are amenable to high-accuracy $N$-body simulations,
which we hope to carry out in the future.

The presence of significant amounts of cold gas in galaxy nuclei
can also accelerate the evolution of a binary SMBH.
However it is not clear whether the net effect of gas would
be to increase, or decrease, the mass of a bound
{\it stellar} population around the coalesced binary,
compared with the purely stellar dynamical estimates made here.
On the one hand, gas dynamical torques can lead to rapid 
formation a tightly-bound binary SMBH \citep{Mayer:07},
reducing the time that the binary can deplete the stellar
density in the core on scales of the SMBH influence radii.
On the other hand, the formation of a steep Bahcall-Wolf
cusp around a shrinking binary discussed in \S 3 requires that the
binary evolution timescale be of order the nuclear relaxation 
time.
Cold gas also implies star formation, which could increase the number
of bound stars.

\begin{figure}
\includegraphics[clip,width=0.47\textwidth]{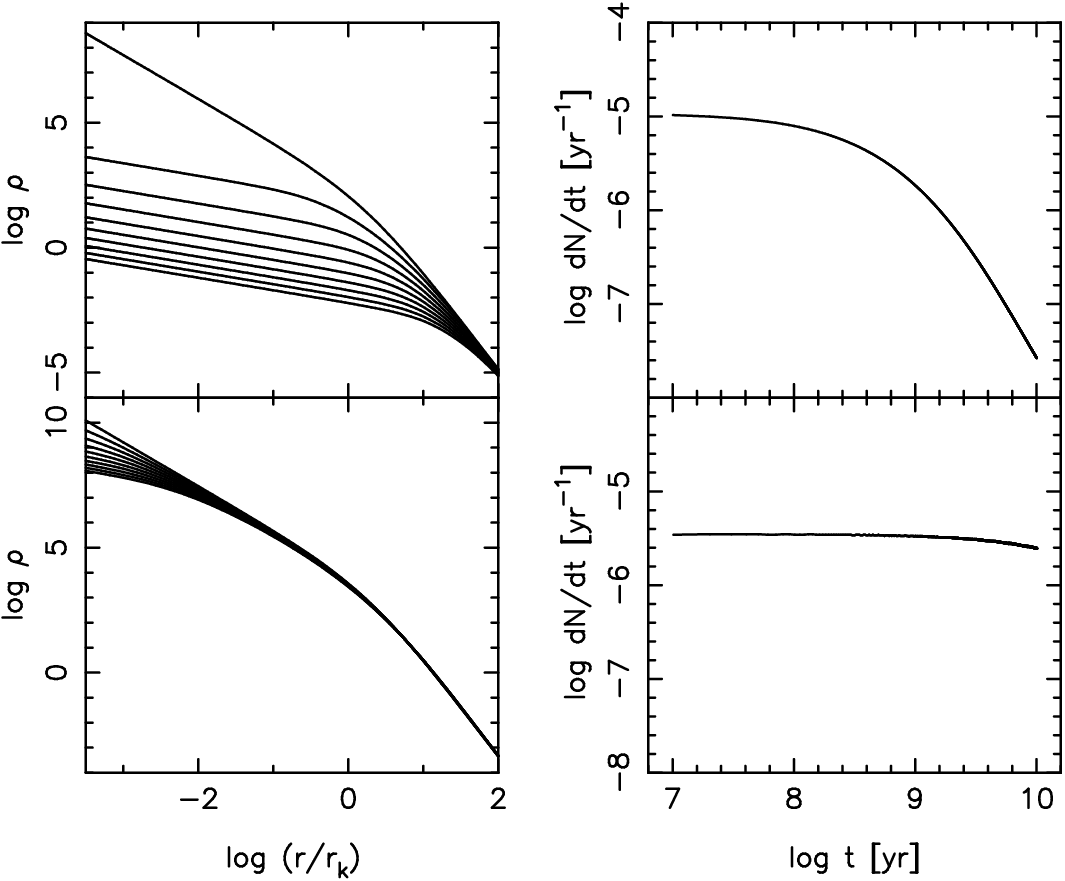}
\caption{Evolution of the density around two \rgcs\ due
to resonant scattering of stars into the SMBH's tidal
disruption sphere.
{\it Top:} $\mh=3\times 10^6\msun$, $\mb\approx 7\times 10^3\msun$;
{\it bottom:} $\mh=3\times 10^7\msun$, $\mb\approx 1\times 10^5\msun$;
$\vk=10^3$ km s$^{-1}$ in both cases.
Left panels show the stellar density at Gyr time increments; 
the density drops as stars are lost into the SMBH.
Right panels show $\dot N$.
}
\label{fig:evol}
\end{figure}

\section{Post-Kick Dynamical Evolution}

In the collisional regime, $\mh\lap 10^7\msun$, a \rgc\ will
continue to evolve via two-body relaxation after it departs the
nucleus.
We argued above that the density profile around the SMBH
will be close to the ``collisionally relaxed'' Bahcall-Wolf form, 
$\rho\propto r^{-7/4}$, at the time of the kick.
After the kick, the Bahcall-Wolf cusp is steeply truncated
at $r\gtrsim \rk$, with $\rk\ll\rinfl$.
Gravitational encounters will continue to drive a flux of stars into
the tidal disruption sphere of the recoiling SMBH, but because 
there is no longer a source of stars at $r\approx\rinfl$ to replace 
those that are being lost, the density at $r\lap\rk$ will steadily drop,
at a rate that is determined by the tidal destruction rate.
The latter is roughly (Paper I)
\beq
\dot N \approx  
{\ln\Lambda\over\ln(\rk/r_{\rm t})}
\left({\vk\over\rk}\right) f_{\rm b}
\label{eq:RR}
\eeq
stars per unit time, where $r_{\rm t}$ is the tidal disruption radius.
Equation~(\ref{eq:RR}) is the so-called ``resonant relaxation''
loss rate \citep{RT:96} and it differs by a factor $\sim f_{\rm b}^{-1}$ 
from the standard, non-resonant relaxation rate.
In the case of SMBHs embedded in nuclei, most of the 
disrupted stars come from radii $r\approx\rinfl$ where
resonant relaxation is not effective; once these stars have
been removed by the kick, the stars that remain are almost all
in the resonant regime and equation~(\ref{eq:RR}) is appropriate.

Using equation~(\ref{eq:RR}), the condition that significant
loss of stars take place in $10^9$ yr or less, i.e.
\beq
\left|{1\over N} {dN\over dt}\right|^{-1} \lap 10^9 {\rm yr},
\eeq
becomes
\beq
\left({\vk\over 10^3 {\rm km\ s}^{-1}}\right)^{3/2}\gap 
{\mh\over 10^7\msun}.
\eeq
This can be recast as a relation between $\mb$ and $r_{\rm eff}$
using the equations in the previous sections; the resulting
line is plotted as the magenta dot-dashed curve in Figure~\ref{fig:RM}.
\rgcs\ to the left of this line (in the collisional regime only)
should expand appreciably on Gyr timescales.

To simulate the evolution of a \rgc\ in this regime,
we solved the orbit-averaged isotropic Fokker-Planck equation for
stars moving in the point-mass potential of a SMBH.
In its standard form, based on the non-resonant
angular-momentum diffusion coefficients, the Fokker-Planck 
equation would predict evolution rates that are
orders of magnitude too small. 
Instead we used an approximate resonant diffusion coefficient 
as in Hopman \& Alexander (2006).
The amplitude of this diffusion coefficient is not
known from first principles and we chose it to 
approximately reproduce the $N$-body diffusion rates observed 
in Paper I and Harfst et al. (2008).
Unlike in most applications of the Fokker-Planck equation,
the outer boundary condition in our case is $f(E=0)=0$, 
i.e. the density of stars falls to zero far from the SMBH
(in addition to being zero near the tidal disruption sphere).

Figure~\ref{fig:evol} shows the evolution over $10$ Gyr 
for two \rgcs\ with $\mh=(3\times 10^6, 3\times 10^7)\msun$ 
and $\vk=10^3$ km s$^{-1}$.
The first cluster lies to the left of the magenta dot-dashed line in
Figure~\ref{fig:RM} and the second lies to the right.
Tidal disruption rates are initially similar for the two clusters,
$10^{-6}{\rm yr}^{-1}\lap \dot N\lap 10^{-5}{\rm yr}^{-1}$,
but the resultant density evolution is much greater in the smaller
\rgc\ since its initial (stellar) mass ($\sim 10^4\msun$) is less.
The stellar disruption rates in Figure~\ref{fig:evol}, and their change
with time, are similar to estimates made in a simpler way by 
Paper I.

The expansion seen in Figure~\ref{fig:evol} is only significant
for \rgcs\ that are older than a few Gyr and that populate the
leftmost part of the mass-radius plane  (Figure~\ref{fig:RM}).
Furthermore, the theory of resonant-relaxation-driven evolution
of star clusters is still in a fairly primitive form and
the true evolution is likely to be affected in important ways
by mass segregation and other effects that have so far hardly
been studied.
For these reasons we chose to ignore the expansion in what
follows; we note here only that the lowest-mass \rgcs\ are
most likely to be affected by the expansion.
We hope to return to this topic in more detail in later papers.

\begin{figure}
\includegraphics[clip,width=0.45\textwidth]{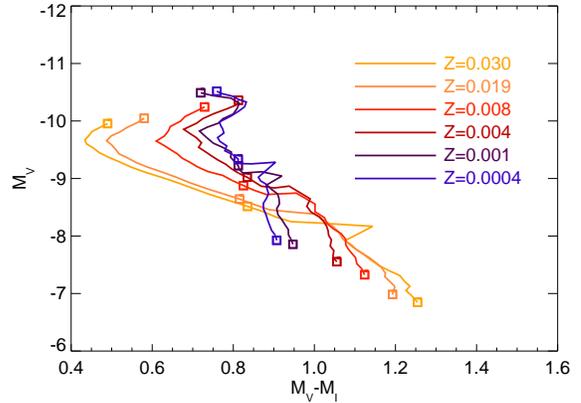}
\caption{Color-magnitude diagram of stellar cluster evolution tracks	
for a range of metallicities and total stellar mass of 
$10^5 M_\odot$. The clusters evolve from the upper-left to 
the lower-right, with open squares corresponding to ages of $10^8$,
$10^9$, and $10^{10}$ years.
}
\label{fig:color_mag}
\end{figure}

\section{Luminosities and Colors}\label{lum_evolution}

Given the total stellar mass in the cluster, we calculate the
luminosity and color of the object using the tabulated stellar
evolution tracks of Girardi et al.\ (1996). These data give absolute
magnitudes in [U,B,V,R,I,J,H,K] bands for a particular stellar (birth) mass,
age, and metallicity. We assume that all the stars in the cluster have
the same age, and the initial mass function (IMF) is given by a
broken power-law distribution \citep{Kroupa:01}:

\begin{equation}\label{eq:imf}
\phi(M) \propto \left\{ \begin{array}{lcc}
    M^{-0.3} & : & 0.08M_\odot > M \\
    M^{-1.3} & : & 0.5 M_\odot > M > 0.08M_\odot \\
    M^{-2.3} & : & M > 0.5M_\odot \end{array} \right.
\end{equation}

At any given time, the total luminosity of a stellar cluster is
generally dominated by red giants, yet the additional contributions from
the main sequence stars shifts the clusters bluer with respect to
individual giants. For example, this means that for a given $B-K$
color, a cluster will have a smaller value of $B-V$ color than an
individual giant with the same $B-K$. Thus unresolved HCSSs may be initially
distinguished from foreground stars by doing a simple cut in
color-color space. Figure \ref{fig:color_mag} shows stellar cluster
evolution on a color-magnitude plot, with age progressing from
the upper-left ($10^8$ years) to the lower-right ($10^{10}$
years). Here we have fixed the total cluster mass at $10^5
M_\odot$, comparable to a typical GC.

If the HCSSs have higher metallicities than typical GCs (a reasonable
assumption if the HCSSs are simply displaced nuclei), we expect them
to have a significantly wider range of colors, which may be useful in
selecting target objects photometrically from a wide field of view. In
particular, since the recoiled star clusters are expected to be quite
old, they should be particularly red compared to GCs of similar ages.

Given the mass-luminosity relation of any individual HCSS, we can now estimate
the luminosity distribution function for a large number of sources.
To arrive at an observed source count, we must first begin
by calculating the formation rates of HCSSs via SMBH mergers. Since
the lifetime of HCSSs is essentially the Hubble time, we need to
integrate the cosmological merger history of the universe beginning at large
$z$ ($\gtrsim 8$) up until today. While these merger rates are
uncertain within at least an order of magnitude, most estimates share
the same qualitative behavior, with the merger rates as observed today
peaking around redshift $z\approx 2-3$
\citep{Menou:01,Sesana:04,Rhook:05} and totalling $\sim 10$ mergers
per year (as measured by an observer at $z=0$) for $M_{\rm BH} > 10^5
M_\odot$.

We follow the results of Sesana et al. (2004) in estimating merger
rates as a function of total black hole mass and redshift. In practice, this
entails defining an ad-hoc mass distribution function of merging SMBHs
with the form
\begin{equation}
\Phi(M,z) \sim f(M)g(z),
\label{eq:phimz}
\end{equation}
where $f(M)$ and $g(z)$ are constructed to match the results of Figure
1 from Sesana et al. (2004). They find that, at any given redshift, the
merger rate scales roughly as $M^{-3/2}$ per log mass. This corresponds
to a functional form of $f(M)\sim M^{-5/4}$, and $g(z)$ is well-described
by a polynomial with an exponential cutoff at large $z$. Then the rate
of observed mergers with total mass $M=M_1+M_2$ is given by
\begin{equation}
R(M,z) = \int \Phi(M_1,z)\Phi(M_2,z)dM_1 .
\label{eq:rmz}
\end{equation}

\begin{figure}
\includegraphics[clip,width=0.45\textwidth]{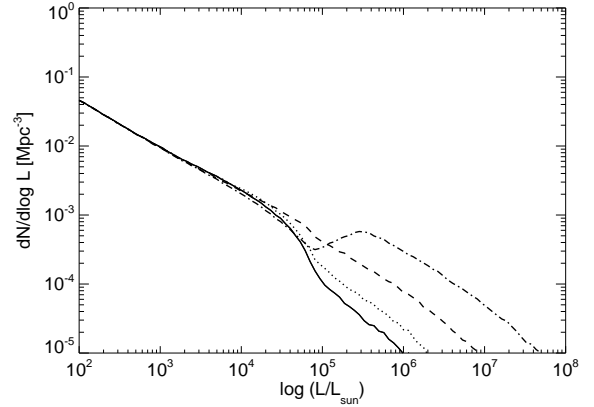}
\caption{Luminosity distribution function of HCSSs per comoving
  Mpc$^3$. For the collisionless case with $M_{\rm BH} \gtrsim 10^7
  M_\odot$, the curves correspond to $\gamma = 0.5, 1.0, 1.5, 2.0$:
  ({\it solid, dotted, dashed, dot-dashed}).
}
\label{fig:rates_L}
\end{figure}

\begin{figure}
\includegraphics[clip,width=0.45\textwidth]{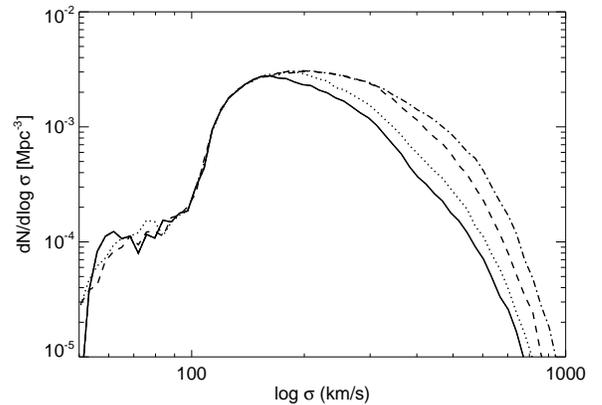}
\caption{Distribution of observed velocity dispersions $\sigma_{\rm
    obs}$ for HCSSs with total luminosity $L > 10^4
    L_\odot$. For the collisionless case with $M_{\rm BH} \gtrsim 10^7
  M_\odot$, the curves correspond to $\gamma = 0.5, 1.0, 1.5, 2.0$:
  ({\it solid, dotted, dashed, dot-dashed}).
}
\label{fig:rates_sig}
\end{figure}

\begin{figure}
\includegraphics[clip,width=0.45\textwidth,angle=-90.]{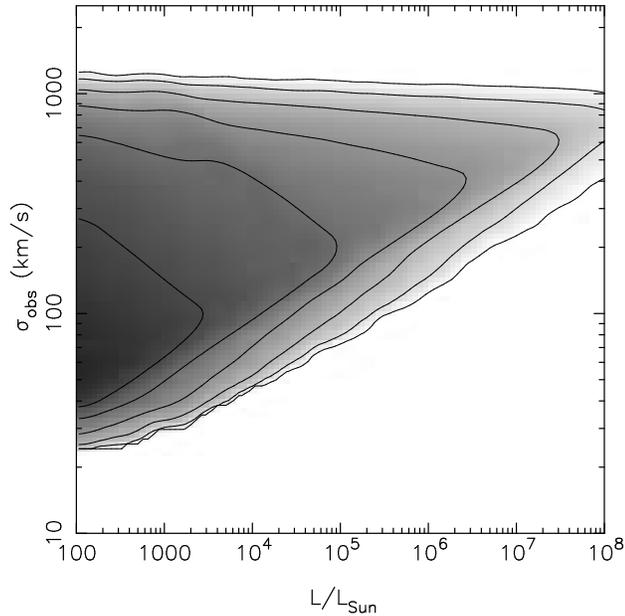}
\caption{Number density of HCSSs with a given luminosity and observed
    velocity dispersion, in units of $d^2N/(d\log L\, d\log \sigma)$
    per Mpc$^3$, for an age at formation of $10^8$ years. The contour
    lines (from left to right) correspond to values of $\log N =
    (-2,-3,-4,-5,-6,-7)$. 
}
\label{fig:contour_Ls}
\end{figure}

To model the merger history of the universe, we follow the same
approach as in Schnittman \& Krolik (2008), integrating forward in time
from redshift $z=8$ (using a standard
$\Lambda$CDM cosmology with $\Omega_{\rm m}=0.3$, $\Omega_\Lambda=0.7$, and
$h=0.72$), and at each redshift, generate a Monte Carlo sample of
merger pairs, each weighted appropriately from the distribution
function $\Phi(M,z)$. We
then normalize the total merger rates to reproduce the results of
Sesana et al.\ (2004), as observed today. 
In the Monte Carlo sampling, we constrain the selected SMBHs to have a mass
ratio $q$ greater than $10^{-3}$, motivated by the
dynamical friction
timescale for the tidal stripping of the satellite to be less than a
Hubble time. In any case, the rate at which HCSSs are ejected from the
galaxy is not dependent on the precise value of the mass-ratio cutoff,
since at mass ratios smaller than $\sim 0.1$, there is little
appreciable kick. 

For each merger, if the resulting SMBH recoil is large enough to escape
from the host galaxy, we consider it to form a HCSS. For a given mass
ratio, the kick velocity is calculated using
equations (1--4) from Baker et al. (2008), assuming spin magnitudes
in the range $0.5\le a_{\rm 1,2}/M_{\rm 1,2} \le 1.0$, and spin orientations with a random uniform
distribution. These assumptions are reasonable if SMBHs gain most of
their mass through accretion (thus a relatively large spin parameter)
and come together through dynamical friction after their host galaxies
merge (thus random spin orientations). As pointed out by
Bogdanovic et al. (2007), gas-rich or ``wet'' mergers may result in
rapid alignment of the two SMBH spins, producing significantly
smaller recoils. On the other hand, ``dry'' mergers should allow the
SMBHs to retain their original random orientations
\citep{Schnittman:04}. However, even in wet mergers, a circumbinary disk may
form and drive the two SMBHs together via gas-dynamical torque without very much direct accretion
onto either SMBH, and therefore remain relatively dry, with
correspondingly large kicks.

\begin{figure}
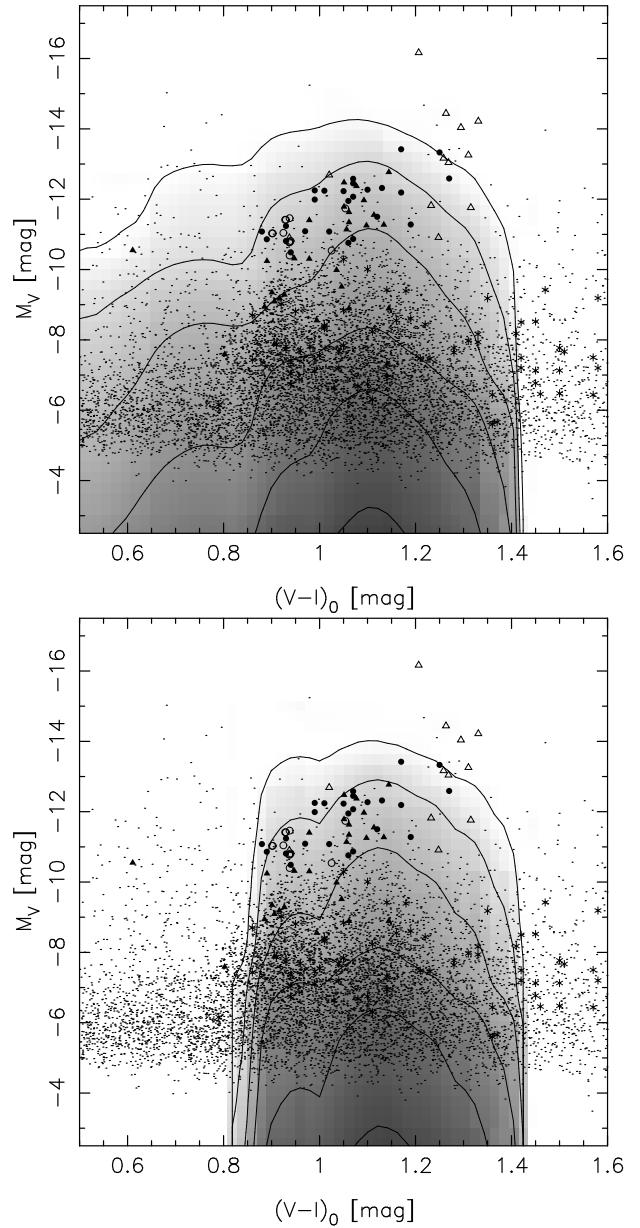

\includegraphics[clip,width=0.45\textwidth,angle=-90.]{fig_ColorvsMag_VI_1e8.ps}
\includegraphics[clip,width=0.45\textwidth,angle=-90.]{fig_ColorvsMag_VI_5e9.ps}
\caption{Number counts of HCSSs with a given $(V-I)$ color
    and absolute visual magnitude $M_V$,
    in units of $d^2N/(dM_V\, d(V-I))$ per Mpc$^3$, with the same
    shading and contour line values as Fig.\ \ref{fig:contour_Ls}.
    {\it Top}: star formation at $t_{\rm k}-0.1$ Gyr;
    {\it bottom}: star formation at $t_{\rm k}-5$ Gyr.
    Filled circles are UCD's from Evstigneeva et al. (2008).
    Open circles are DGTO's from Hasegan et al. (2005).
    Triangles are E-galaxy nuclei from Cote et al. (2006);
    open triangles are nuclei brighter than $B_T=13.5$ and 
    filled triangles are nuclei fainter than $B_T=13.5$.
    Stars are Milky Way GCs from Harris (1996) and points are
    Virgo cluster GCs from Mieske et al. (2006).
    }
\label{fig:contour_CM}
\end{figure}

\begin{figure}
\includegraphics[clip,width=0.47\textwidth,angle=-90.]{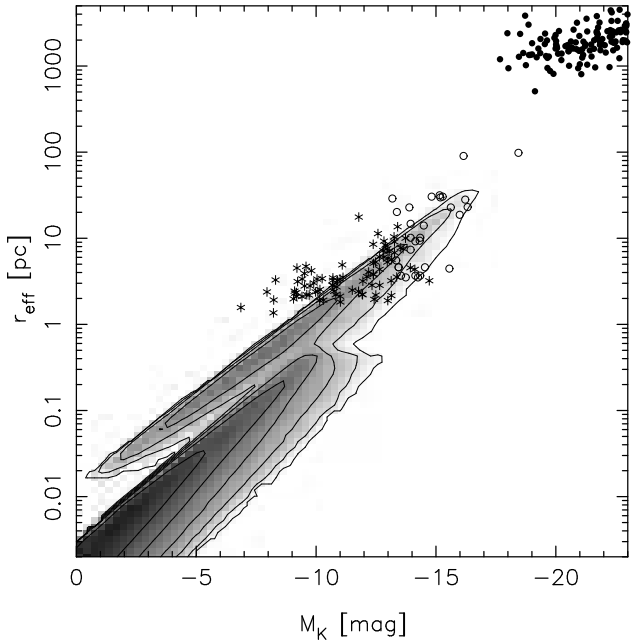}
\caption{
    Number counts of HCSSs with a given effective radius $r_{\rm eff}$
    and absolute $K$-magnitude $M_K$,
    in units of $d^2N/(d\log r_{\rm eff}\, dM_K)$ per Mpc$^3$ .
    Contour levels are the same as in Fig.~\ref{fig:contour_Ls}.
    Data points are from Forbes et al. (2008).
    Filled circles: E galaxies.
    Open circles: UCDs and DGTOs.
    Stars: globular clusters.
    }
\label{fig:contour_RM}
\end{figure}

\begin{figure}
\includegraphics[clip,width=0.47\textwidth,angle=-90.]{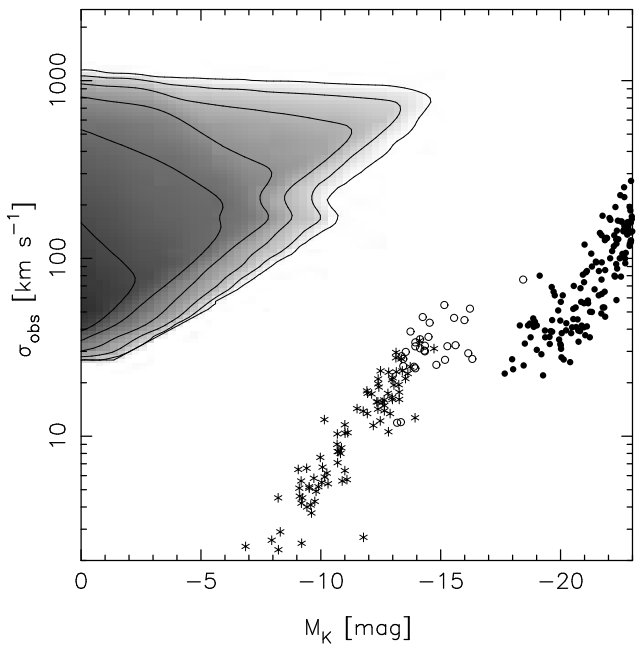}
\caption{
    Number counts of HCSSs with a given velocity dispersion 
    $\sigma_{\rm obs}$ and absolute $K$-magnitude $M_K$,
    in units of $d^2N/(d\log \sigma_{\rm obs}\, dM_K)$ per Mpc$^3$ .
    Contour levels and data points are the same as in 
    Fig.~\ref{fig:contour_Ls}.
}
\label{fig:contour_SM}
\end{figure}\section{Rates of Production}

From Figure \ref{fig:escape}, we estimate that the escape velocity is
roughly five times the nuclear stellar velocity dispersion $\sigma$, 
where $\sigma$ is
determined from the total SMBH mass from equation (\ref{eq:MS})
above. If the final SMBH does escape, it will carry along a total mass
$M_{\rm b}$ in bound stars, determined by equations
(\ref{eq:Mb_collisional}) for collisional relaxation 
($M_{\rm BH} \lesssim 10^7 M_\odot$) and
(\ref{eq:Mb_collisionless}) for collisionless relaxation ($M_{\rm BH}
\gtrsim 10^7 M_\odot$). 
Neglecting mass loss from stellar winds and tidal disruptions, 
the total cluster mass in stars should remain roughly constant 
over a Hubble time. 
We used the stellar evolution tables from Girardi et al. (1996) 
to calculate the colors and luminosity 
of each ejected HCSS for two cluster ages: (1) the stars in the cluster were
formed $10^8$ years before SMBH merger and ejection, and (2) the stars
formed $5\times 10^9$ years before ejection.
We carried out this exercise both with a single (solar)
metallicity for all \rgcs, 
and also assuming a Gaussian distribution of metallicities:
\begin{subequations}
\begin{eqnarray}
N(x) &=& {1\over\sqrt{2\pi}\sigma} \exp\left[-\left(x-\overline{x}\right)^2/2\sigma^2\right],\\
x&\equiv& \log_{10}\left[{Z_i\over Z_\odot}\right].
\end{eqnarray}
\label{eq:nofx}
\end{subequations}
We took
\begin{equation}
\overline{x} = -0.5, \ \ \ \ \sigma = 0.5
\end{equation}
which correspond approximately to what Cot\'e et al. (2006)
inferred as the metallicity distribution of nuclear star
clusters in the Virgo cluster galaxies, assuming ages of 5 Gyr.
We obtained very similar results under the two assumptions,
in part because the Girardi et al.\ (1996) tables only extend to
slightly  super-solar metallicities ($Z=0.03$).
Figures~\ref{fig:rates_L}--\ref{fig:contour_Ls}
assume solar metallicity while Figures~\ref{fig:contour_CM}--
\ref{fig:contour_SM} assume the metallicity distribution
of equation~(\ref{eq:nofx}).

Figure \ref{fig:rates_L} shows the luminosity distribution function
(number per comoving Mpc$^3$) for a range of $\gamma$, with a stellar
age of $10^8$ yr at time of SMBH recoil. The
high-luminosity systems correspond to high-mass SMBHs that merge via
collisionless relaxation and thus the number of bound stars is
directly a function of the parameter $\gamma$. In Figure
\ref{fig:rates_sig} we plot the distribution as a function of
observed velocity dispersion, limited only to systems with $L > 10^4 L_\odot$. The low-velocity cutoff is directly a
function of the minimum SMBH mass needed to keep roughly $10^4 M_\odot$
in bound stars; any galactic
halo with SMBH mass above $\sim 10^6 M_\odot$ will have an escape
velocity $\gtrsim 400$ km/s, and thus an observed velocity dispersion
$\sigma_{\rm obs} \gtrsim 120$ km/s. The high-velocity cutoff is
defined by the maximum kick velocity of $V_{\rm kick} \lesssim 4000$
km/s, as determined by numerical simulations of BH mergers. While the
majority of HCSSs will have small velocity
dispersions, the brightest, most massive ones will likely come from
the most massive host galaxies, and thus require the largest kicks, in
turn giving the highest internal velocity dispersions. In this regard,
HCSSs behave similarly to more classical stellar systems: higher
masses have higher dispersion.
However, holding the SMBH mass fixed, a larger recoil velocity will
result in a {\it smaller} number of bound stars
[eqns. (\ref{eq:Mb_collisional}, \ref{eq:Mb_collisionless})], and thus
{\it lower} luminosity for higher velocity dispersion. 

In Figure \ref{fig:contour_Ls} we show a contour plot of the density
of HCSSs as a function of luminosity and observed velocity
dispersion. Here we clearly see that the most luminous systems will
also have the largest dispersion, roughly an order of magnitude
greater than any globular cluster of the same stellar mass. We can
also use Figure \ref{fig:contour_Ls} to estimate the number of HCSSs
that might be observable in the local universe. Assuming a uniform
spatial distribution in the local universe, out to a
distance of 20 Mpc ($\sim$30,000 Mpc$^3$), for $\gamma=1$ we should expect to see dozens
of objects with $L>10^4L_\odot$ and at least a few with
$L>10^5L_\odot$. However, an all-sky survey to find these few innocuous
objects could be prohibitively expensive. Coincidentally, the total
mass in the Virgo cluster out to a radius of $\sim 2$ Mpc is roughly
the same as that of a smooth universe out to $\sim 20$ Mpc, which is
approximately the distance to Virgo \citep{Fouque:01}. In other words,
a focused survey
of Virgo would be able to sample an effective volume of $\sim 30,000$
Mpc$^3$ all at the same distance and with a relatively small field of
view. 

We expect to find 
$\sim 1$ \rgc\ in Virgo with $m_K \le 20$; $\sim 6$ with $m_K \le 22$;
$\sim 40$ with $m_K \le 24$; and $\sim 150$ with $m_K \le 26$, almost
all of which would have
$\sigma_{\rm obs} \gtrsim 200$ km s$^{-1}$. For the Fornax cluster,
which is at roughly the same distance, but contains less mass by a
factor of $\sim 15$ \citep{Drinkwater:01}, the source counts at the
same fluxes should be down by a comparable factor. The Coma cluster,
on the other hand, has a mass comparable to Virgo \citep{Kubo:07}, but
at a distance of $\sim 100$ Mpc, the apparent brightness of any HCSSs
will be smaller by $\sim 4$ magnitudes.

Figures~\ref{fig:contour_CM},
\ref{fig:contour_RM} and \ref{fig:contour_SM} show predicted number
counts in the color-magnitude, size-magnitude, and velocity 
dispersion-magnitude planes.
Over-plotted for comparison are data for other compact stellar systems,
from various sources, as discussed in the figure captions.
In constructing these figures (as well as Figure~\ref{fig:contour_Ls}), 
we set $\gamma=1$ in the ``collisionless'' regime.
This resulted in a slight bimodality in the distributions corresponding
to the discontinuous change in $\gamma$ from $1.75$ to $1$ at galaxy 
masses of $\sim 10^{11}M_\odot$.
Alternatively, one could allow $\gamma$ to vary in some smooth way
with galaxy mass (cf. the discussion in  \S3.2).
 
\section{Identifying \rgcs}
\subsection{Search Strategies}

We have shown that stars bound to a recoiling SMBH would
appear as very compact stellar clusters
with exceptionally high velocity dispersions.
The density of \rgcs, and therefore
the chances of finding them, will be highest in
clusters of galaxies, and nearby galaxy clusters
like Virgo and Fornax are therefore well suited to 
searching for \rgcs.

In their properties, HCSSs share similarities with globular
clusters (GCs). However, they would differ from classical
GCs by their much larger velocity dispersions
and (possibly) higher metal abundances
(the nuclei of elliptical galaxies in the Virgo and Coma
clusters often have metal abundances comparable with solar,
and the cores of quasars, powered by major mergers,
frequently show super-solar metallicities).
They would differ from stripped galactic nuclei and ultracompact
dwarf galaxies (UCDs, Phillips et al. 2001) by their typically
greater compactness (Figure~\ref{fig:contour_RM}).
They would differ from objects in their {\em local} environment by
possibly showing a large velocity offset.

How can we find HCSSs and distinguish them from other
source populations?
Systematic searches would be based on color, compactness,
spectral properties, or combinations of these.

Imaging searches for compact stellar systems have been or 
are currently being carried out, focussing especially
on the nearest clusters of galaxies Fornax 
\citep[e.g.][]{Hilker:99,Drink:03,Mieske:08},
Virgo \citep[e.g.][]{ACSI,Hasegan:05,Mieske:06,Firth:08},
and Coma \citep{Carter:08},
and on a number of nearby groups \citep{Evst:07}.
These studies have resulted in the detection of a large number
of GCs, and of several UCDs per galaxy cluster.

Two strategies suggest themselves for identifying \rgcs\ among
existing surveys of compact stellar systems in nearby galaxy
clusters.
One would focus on the faintest \rgcs\ which are most abundant;
the other would concentrate on the brightest objects,
which are rare, but most amenable to follow-up spectroscopic
observations.

According to Figure~\ref{fig:contour_RM}, \rgcs\ 
separate in $r_{\rm eff}-$luminosity space most strongly at the smallest
effective radii, smaller than Galactic GCs, 
while Figure~\ref{fig:contour_CM} suggests that
their colors should be comparable to (metal rich) GCs or 
(gas-poor, i.e. non-star-forming, and non-accreting) galactic nuclei.
PSF-deconvolved HST imaging
can achieve a spatial resolution better than $0.1$ arcsec,
corresponding to spatial scales of $\sim 10$ pc at the
distance of the Virgo cluster.
However, in order to confirm such \rgc\ candidates, a laborious
spectroscopic multi-fiber follow-up survey would then have
to be carried out.

Instead, selecting brighter (even though rarer) objects appears more
promising and would substantially reduce the exposure time.
A key signature of a \rgc\ is its large velocity dispersion,
which would distinguish it from luminous GCs and most known UCDs.
At the same time, the highest velocity dispersions would tend to
put the broadened absorption lines below the noise.
Therefore, HCSSs with $\sigma_{\rm obs}$ 
 below several hundred km s$^{-1}$ might be easiest to detect.
These are in fact expected to be the most common
(Figure~\ref{fig:rates_sig}).

In order to estimate exposure times, we simulated spectra with
the multi-object spectrograph FLAMES attached to the VLT  
\citep{Pasquini:02}, using the spectrograph GIRAFFE in MEDUSA mode.
This spectrograph allows the observation of up to 130
targets at a time at intermediate ($\sim 30$ km s$^{-1}$)
to high ($\sim 10$ km s$^{-1}$) spectral resolution.
The simulated spectral deconvolutions in \S2 suggest that a signal-to-noise
(S/N) ratio of $\sim 10$ is sufficient to detect the broadened lines, 
while S/N $\approx 30$ is desirable in order to probe the non-Gaussianity
of the broadening function.
%The latter may be desirable since the large velocity dispersion, small
% size, and (possibly) high space velocity of a \rgc\ could mimic a distant
%background galaxy; detection of an extremely non-Gaussian broadening function
%would effectively rule out this possibility.
In order to reach a S/N of 30 for a cluster 
of $m_V\approx 21$ or fainter ($M_V\gap -10$ at Virgo) 
requires excessive integration times in the high 
resolution mode.
In lower resolution mode, about 10 hours exposure time are
required to reach S/N=30 for $m_V=20$.
Simply detecting the high velocity dispersion requires less time,
of order an hour or less for $m_V\lap 21$.

\subsection{Could UCDs be \rgcs?}

\rgcs\ share some properties with UCDs and dwarf-globular
transition objects (``DGTOs''; Hasegan et al. 2005).
These are compact stellar systems with stellar velocity
dispersions as high as $\sim$50 km s$^{-1}$, 
masses between 10$^{6-8}$ $M_{\odot}$
and unusually high mass-to-light ratios \citep[e.g.][]{Hilker:08}.
Figures~\ref{fig:RM} and~\ref{fig:contour_RM} suggest that UCD 
sizes are consistent with those of the largest (``collisionless'') \rgcs, 
although Figures~\ref{fig:SM} and~\ref{fig:contour_SM} suggest that 
known UCD velocity dispersions are too low by a factor of at least a few.
Furthermore Figures~\ref{fig:contour_RM} and ~\ref{fig:contour_SM}
suggest that \rgcs\ with properties similar to those of UCDs are 
likely to be rare.
However, there is increasing evidence that UCDs are a
``mixed bag'' (e.g., Hilker 2006), possibly requiring a 
number of different formatios mechanisms
\citep{Oh:95,FK:02,Bekki:03,Mieske:04,Martini:04,Goerdt:08}.
Individual \rgcs\ might therefore hide among the UCD population,
and low-mass UCD and DGTO candidates identified in current
and future surveys might sometimes be recoiling \rgcs.
Objects like  Z2109 \citep{Zepf:08} with its very unusual
and broad [OIII] emission line are also possible candidates. 

\subsection{Very young \rgcs?}

The predicted \rgc\ colors and luminosities summarized in
Figures~\ref{fig:contour_CM} - \ref{fig:contour_SM} reflect the
properties of the old ($\sim 5$ Gyr) stellar populations that
predominate in our models.
A \rgc\ that was discovered near its birthplace might appear
much younger than implied by these plots,
particularly if it was born in a starburst galaxy resulting
from a merger.
The light from such a \rgc\ might be dominated
by young massive stars, at least during the $\sim 0.1-1$ Gyr required 
for it to exit the galaxy.
(We are assuming here that massive stars can form, or at least
quickly find their way, well inside the SMBH influence radius, as
appears to be the case at the center of the Milky Way.
We are further assuming that the recoiling SMBH is not
accreting at the time of observation since otherwise the quasar
might outshine the starlight.)
We implicitly excluded this possibility above by assuming $0.1$ Gyr as a
minimum lag between star formation and SMBH ejection.

Starting roughly 6 Myr after a starburst episode, 
luminosities and colors of the burst population are
expected to increase rapidly due to red supergiant stars (RSGs), 
which begin to appear when main sequence stars of initial mass 
$\sim 25 M_\odot$ finish core-hydrogen burning.
The absolute magnitude of a burst population with standard initial 
mass function spikes roughly $\sim 10$ Myr after the burst due to the RSGs;
the increase is greatest in the IR bands. For example, in the K band, 
integrated luminosity increases to a value $\sim 100$ times greater than 
for the same population at $1$ Gyr (Leitherer et al. 1999) before
falling off rapidly after $\sim 14$ Myr.
Integrated colors reach their reddest values at the same time,
i.e. V-I $\approx 1.5$.
Single RSG stars can approach luminosities of $10^5- 10^6 L_\odot$
(Davies et al. 2007), so the luminosity and color of an \rgc\ 
containing $\ll 10^6$ stars could undergo enormous relative changes 
around this time.

There are both advantages and disadvantages to targetting such
young \rgcs\ as a search strategy.
(Super)star clusters are commonly found in the centers of
starburst galaxies and merger remnants \citep[e.g.][]{Max:05,Galliano:08},
so in any search based solely on imaging photometry, 
\rgcs\ could not be easily distinguished.
However, once a \rgc\ reaches larger separations from the center
of the galaxy, it will already deviate from its surroundings by 
plausibly showing higher metal abundance.
On the other hand, even low-resolution spectroscopy could immediately 
reveal a high velocity relative to the galaxy core or relative to
other clusters in the galaxy.

We note that discovery of a young \rgc\ that is 
still uniquely associated with its host galaxy can
provide important constraints on the time scale associated with
late evolution of the binary SMBH that engendered the kick.
If coalescence was rapid, the host galaxy of the \rgc\ will
still show signs of recent interaction (for instance
tidal tails). If the massive binary stalled for 
$\gap 10^8-10^9$ yr before coalescing, these signs will be 
mostly gone.
Age dating of the recoil event could be based on (projected) 
distance from the galaxy core and recoil velocity, 
and/or on the age of the stellar populations.

\subsection{Exotic manifestations}
\rgcs\ with {\em accreting} central black holes, powered
either by stellar tidal disruptions or stellar mass loss,
were discussed in Paper I. 
This sub-population could be efficiently searched for by 
combining optical properties with information from multi-wavelength 
surveys, e.g., in the X-ray, UV or radio band.
We speculate that the unusual optical transient source SCP06F6 
\citep{Barbary:08} might be a tidally-detonated white dwarf 
bound to a recoiling SMBH (as discussed already in Paper I). 
This scenario would fit the high amplitude of variability of the 
transient, the absence of an obvious host galaxy, 
and the possible association with a cluster of galaxies at $z=1.1$.  
However, the observed symmetry of  the lightcurve might suggest
a  lensing origin \citep{Barbary:08}.
The unusual optical spectrum of this source could be
caused by the tidal-debris disk illuminating the outer disk
or the outflowing part of the detonation debris. 
That way, the observed extreme, unusally broadened
 absorption features would be caused if we are looking down-stream.
 Gaensicke et al. (2008) recently reported
the detection of an X-ray source co-incident with SCP 06F6
with an X-ray luminosity at the lower end of known
 tidal disruption flares \citep{Komossa:04}.
These authors discuss supernoave-related scenarios
 but also consider tidal disruption of a star, and suggest a preliminary
 redshift of 0.14, in which case the source is not associated with the
 cluster at redshift 1.1.

Finally, some of the oldest surviving \rgcs\ would
consist mostly of stellar end states. They would
be quite faint, since only very low-mass stars and WDs would
remain,  but could possibly be identified
by their very unusual colors.

\section{The Inverse Problem}

We have focused on the ``forward'' problem of predicting
the numbers and properties of \rgcs\ given reasonable assumptions
about the distribution of gravitational wave kicks and
the merger history of the universe.
Once \rgcs\ have been detected, one can begin 
work on the potentially more interesting inverse problem:
using the measured  properties of \rgcs\  to infer the 
distribution of GW kicks and its evolution over time.

The inverse problem is made easier by
the remarkable property of \rgcs\ (\S2) that they encode
the magnitude of their natal kick in their spectra.
Measuring the degree to which the absorption-line spectrum
of a \rgc\ has been broadened by internal stellar motions
leads immediately to an estimate of $\vk$.
Such a measurement is completely independent of the space 
velocity of the \rgc\ at the moment of observation.
It is reasonably independent of the initial (pre-kick) density profile, 
and it depends on the  the time since the kick only to the extent that the
\rgc\ changes its structure over time;
such changes are expected to be small for the brightest \rgcs
(\S 4).

For a ``collisional'' ($\mh\lap 10^7\msun$) \rgc, 
with $\gamma\approx 1.75$, equation~(\ref{eq:sofgamma}) gives
\beq
\ln F_3 = -2.17 + 0.56\times 1.75 ,
\eeq
i.e.
\beq
\vk\approx 3.3 \sigma_{\rm obs}.
\label{eq:vksig}
\eeq
Absent any knowledge about the internal structure of the \rgc,
the coefficient in equation~(\ref{eq:vksig}) is uncertain,
but not greatly so.
Allowing the inner density profile slope to vary over the range
$1\le\gamma\le2$ implies 
\beq
2.9 \lap\vk/\sigma_{\rm obs} \lap 5.0 .
\eeq
If more information about  the \rgc\ is available than 
just $\sigma_{\rm obs}$, this estimate of $\vk$ could be
refined, to a degree that depends on the size and distance 
of the \rgc\ and on the access to observing time:
(1) A deep spectrum would allow extraction of the stellar
broadening function $N(V)$, as in Figure~\ref{fig:mpl}.
$N(V)$ contains more information about the spatial and velocity 
distribution of stars around the SMBH than $\sigma_{\rm obs}$ alone
\citep[e.g.][]{Merritt:93}.
(2) If the \rgc\ is near enough and/or large enough to be spatially resolved,
constraints can be put on the slope of the stellar density profile
from the photometry.

Measuring both $\rk$ and $\vk$ gives 
$\mh$ (eq.~\ref{eq:rk}), allowing one to investigate
the dependence of kick velocity on SMBH mass, and 
(via the $\mh-\sigma$ relation) on galaxy mass.
Combined with the total light of the \rgc\, and perhaps
with a mass-to-light ratio derived from broad-band colors,
$\rk$ and $\mh$ give an estimate of the pre-kick nuclear density
via equation~(\ref{eq:f1}).

Most detected \rgcs\ may be spatially unresolved.
Even in this case, broad-band magnitudes would allow a sample
of \rgcs\ to be placed on the color-magnitude or
velocity dispersion-magnitude diagrams 
(Figs.~\ref{fig:contour_RM}, \ref{fig:contour_SM}).
The number of detected \rgcs\ per unit volume combined
with their distribution over these observational planes 
contains information about the time-integrated ejection rate,
hence the galaxy merger rate.
Colors  would  also provide an indirect constraint on the
time since the kick.

So far we have emphasized kicks large enough to unbind SMBHs from 
galaxies, $\vk\gap 500$ km s$^{-1}$ \citep{Merritt:04}.
If these are the only objects detected as \rgcs\,
then they will contain information only about  the high-$\vk$ 
part of the kick distribution (although we note that a large
fraction of kicks may be above 500 km s$^{-1}$).
Many kicks will fall below galactic escape velocities, 
particularly in the largest galaxies, producing \rgcs\ that oscillate
about the core or drift for long times in the envelope \citep{MQ:04,GM:08}.
Since the size of a \rgc\ scales inversely with its kick 
(eq.~\ref{eq:rk}), such objects would be among the largest 
and brightest \rgcs, but detection might be difficult 
since they would be superposed on or behind the image of the galaxy.
Such \rgcs\ would also have finite lifetimes before
finding their way back to the center of the galaxy.

\section{Conclusions}

1. Supermassive black holes (SMBHs) kicked out from the centers of
galaxies by gravitational wave recoil are accompanied by a cluster of
bound stars with mass $\sim 10^{-2}$ times the black hole mass or less, 
and  radius $\sim 10^1$ pc or less -- a ``hyper-compact stellar system'' 
(\rgc).

2. \rgcs\ have density profiles that can uniquely be
calculated given the kick velocity and given the stellar
distribution prior to the kick.
The density at large distances from the SMBH falls
off as $\sim r^{-4}$.

3. Internal (rms) velocities of \rgcs\ are very high,
$\sim 10^2-10^3$ \kms, and comparable to their kick velocities.
Their overall velocity distributions are extremely non-Gaussian.

4. HCSSs could be distinguished photometrically from foreground red
giants, based on their bluer (lower) values of $B-V$ for a given
$B-K$. They also should appear redder than low-metallicity GCs with
comparable ages $\gtrsim 1$ Gyr.

5. With a simplified cosmological merger model, we are able to
estimate expected number counts and luminosity distributions of HCSSs
in the local universe. Detection of perhaps $10^2$ HCSSs should be
possible in the Virgo cluster alone, although only a few may be bright
enough to allow high S/N spectroscopy and provide solid confirmation
of their extreme velocity dispersions. 

6. Some \rgcs\ may already exist in survey data of compact
stellar sysetms in the Fornax, Virgo and Coma galaxy clusters.

7. Because the kick velocity of a \rgc\ is related in a simple way to
its measured velocity dispersion, 
the distribution of gravitational wave kicks can be empirically determined
from a sufficiently large sample of \rgcs.

Paper I \citep{KM:08a} first derived the basic properties
of \rgcs\, including their compactness and high internal velocity 
dispersions, and presented a possible route to detection via 
off-nuclear tidal disruption flares.
The current paper derives the intrinsic properties of \rgcs\ in a more 
complete and general way and relates those properties to the properties 
of the host galaxy.
Together, these two papers complement the growing number of recent 
papers that discuss gas-related signatures of gravitational wave recoil.
While this paper was being submitted, we learned of a related work by 
O'Leary \& Loeb (2009) which argues that many thousands of low-mass \rgcs\ 
should be present in the Milky Way halo.

\bigskip\bigskip
\acknowledgements
This work was begun while J. S. was a vistor to the Center for
Computational Relativity and Gravitation at the Rochester 
Institute of Technology.
We thank S. Mieske and P. Lasky for making unpublished data 
available to us.
We acknowledge stimulating conversations with D. Axon, A. Gualandris, 
J. Krolik, M. Mbonye, C. Miller, and S. Portegies Zwart. 
The anonymous referee also made insightful comments that led to
the new section on unbound stars.
D. M. was supported by grants AST-0807910 (NSF) and NNX07AH15G (NASA).
J. S. was supported by the Chandra Postdoctoral Fellowship Program.

\appendix
\begin{center}
{\bf A. Computation of the Bound Mass}
\end{center}

Here we compute the mass in stars that remains bound to a SMBH
after the latter receives an instantaneous kick of magnitude $\vk$.

We assume that the pre-kick phase space density of stars is 
\beq
f_0(E) = C\left|2E\right|^{\gamma-3/2}, \ \ \ \ -\infty\le 2E\le 0 \,
\eeq
where $E=V^2/2-G\mh/r$ is the energy per unit mass of a star.
The pre-kick stellar density is $\rho_0(r) \propto r^{-\gamma}$.
As above, we ignore the contribution to the gravitational potential
from the stars.

Immediately after the kick, transfer to a frame moving with the SMBH.
Assume without loss of generality that the kick is along the $x$ axis.
In this frame, the phase space density is
\begin{equation}
f = C\left[{2G\mh \over r} - \left(V_x-\vk\right)^2 - V_y^2 - V_z^2\right]^{\gamma-3/2}
\end{equation}
in the velocity-space region that lies within the spere
\begin{equation}
\left(V_x-\vk\right)^2 + V_y^2 + V_z^2 = 2{G\mh\over r}
\label{eq:sphere2}
\end{equation}
and zero elsewhere.

\begin{figure}
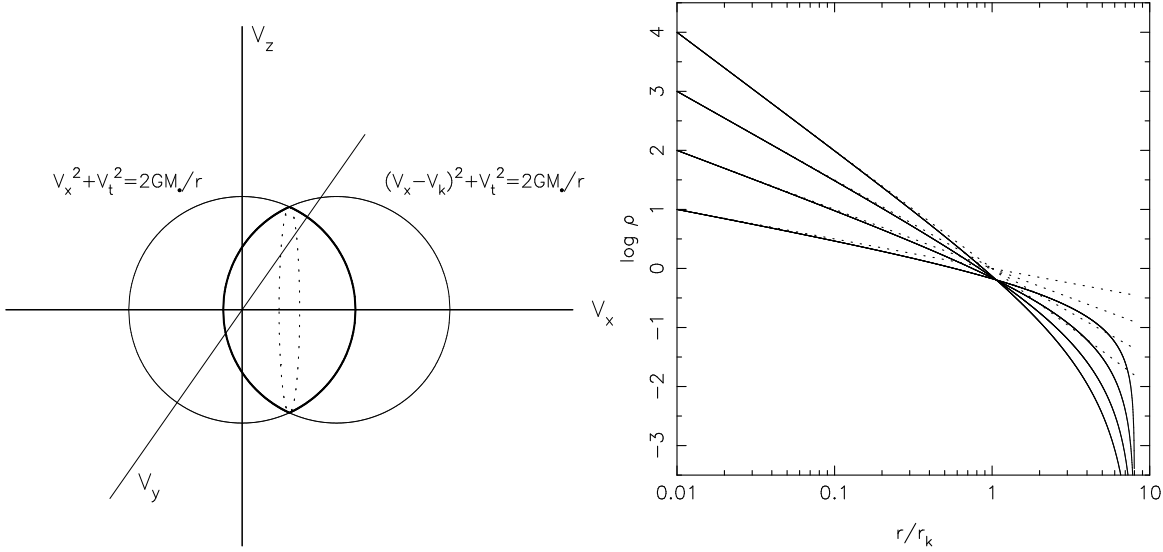

\includegraphics[clip,width=0.45\textwidth,angle=0.]{fig_append1.ps}
\includegraphics[clip,width=0.40\textwidth,angle=0.]{fig_append2.ps}
\caption{(a)
  Illustrating the velocity-space region occupied by
  stars at radius $r$ around a kicked SMBH, in a frame moving
  with the kick velocity.
  Stars that remain bound after the kick lie in the region of
  intersection of the two spheres.
  Solid lines show the density of bound stars around a kicked SMBH, 
  immediately after the kick, for four different values of the pre-kick
  density slope, $\gamma = (0.5,1,1.5,2)$ (dotted lines).
  (b) Solid lines show the density of bound stars around a kicked SMBH, 
  immediately after the kick, for four different values of the pre-kick
  density slope, $\gamma = (0.5,1,1.5,2)$ (dotted lines).
}
\label{fig:append}
\end{figure}

Stars are bound to the SMBH after the kick if they lie within the 
sphere
\begin{equation}
V_x^2 + V_y^2 + V_z^2 = 2{G\mh\over r}.
\label{eq:sphere1}
\end{equation}
The intersection of the surface of this sphere, and the surface of the 
sphere defined by equation~(\ref{eq:sphere2}),  defines a circle of radius
$\left[2G\mh/r-\left(\vk/2\right)^2\right]^{1/2}$
with center on the $V_x$-axis at $V_x=\vk/2\equiv V_0$ (Figure~\ref{fig:append}a).

The configuration-space density of the bound stars is then
\begin{equation}
\rho_{\rm k}(r) = 2\pi C \int dV_x \int dV_t\ V_t \left[{2G\mh\over r} - 
\left(V_x-\vk\right)^2 - V_t^2\right]^{\gamma-3/2}\, ,
\end{equation}
where the velocity-space volume element has been written
$d^3V = 2\pi dV_x V_t dV_t$, $V_t^2\equiv V_y^2 + V_z^2$,
and the region of integration extends over the volume enclosed
by the intersection of the two spheres in Figure~\ref{fig:append}a.

Define new variables
\begin{equation}
x = {V_x-\vk\over \sqrt{2G\mh/r}},\ \ \ \ y = {V_t^2\over 2G\mh/r}.
\end{equation}
The contribution to $\rho_{\rm k}(r)$ from the velocity-space region 
to the left of the dotted curve in Figure~\ref{fig:append}a
(i.e. $V_x\le V_0$) is
\begin{subequations}
\begin{eqnarray}
\rho_I(r) &=& \pi C \left({2G\mh\over r}\right)^\gamma
\int_{-1}^{-\sqrt{r/8\rk}} dx \int _0^{1-x^2} dy 
\left(1-x^2-y\right)^{\gamma-3/2} \\
&=& {\pi C\over \gamma-1/2} \left({2G\mh\over r}\right)^\gamma
\int_{\sqrt{r/8\rk}}^1 dx \left(1-x^2\right)^{\gamma-1/2}\, , 
\end{eqnarray}
\end{subequations}
where $\rk\equiv G\mh/\vk^2$ as above; $\rho_I=0$ for $r>8\rk$.
The contribution to $\rho_{\rm k}(r)$ from the velocity-space region 
to the right of the dotted curve in Figure~\ref{fig:append}a
(i.e. $V_x> V_0$) is
\begin{subequations}
\begin{eqnarray}
\rho_{II}(r) &=& \pi C \left({2G\mh\over r}\right)^\gamma
\int_{-\sqrt{r/8\rk}}^{1-\sqrt{r/2\rk}} dx 
\int_0^{1-x^2-r/2\rk-2x\sqrt{r/2\rk}} dy 
\left(1-x^2-y\right)^{\gamma-3/2} \\
&=& {\pi C\over \gamma-1/2} \left({2G\mh\over r}\right)^\gamma
\int_{-\sqrt{r/8\rk}}^{1-\sqrt{r/2\rk}} dx 
\left[\left(1-x^2\right)^{\gamma-1/2} - 
\left({2r\over\rk}\right)^{\left(\gamma-1/2\right)/2} 
\left(\sqrt{r\over 8\rk} + x\right)^{\gamma-1/2}\right].
\end{eqnarray}
\end{subequations}

Summing $\rho_I(r)$ and $\rho_{II}(r)$, and fixing $C$ by
the requirement that the post-kick density in the limit $\vk\rightarrow 0$ 
equal the pre-kick density $\rho_0(r)$:
\begin{equation}
\rho_0(r) = {3-\gamma\over 2\pi} {\mh\over \rh^3} \left({r\over\rh}\right)^{-\gamma},
\end{equation}
yields
\begin{subequations}
\begin{eqnarray}
\rho_{\rm k}(r) &=& {3-\gamma\over 2\pi I_1(\gamma,0)}{\mh\over\rh^3}
\left({r\over\rh}\right)^{-\gamma}\left[I_1\left(\gamma,r\right) -
\left({2r\over\rk}\right)^{\left(\gamma-1/2\right)/2} I_2\left(\gamma,r\right)\right], \ \ \ \ r \le 8\rk \label{eq:denbounda} \\
I_1\left(\gamma,r\right)&=&\int_{\sqrt{r/2\rk}-1}^1 
dx\left(1-x^2\right)^{\gamma-1/2}, \\
I_2\left(\gamma,r\right) &=& \int_{-\sqrt{r/8\rk}}^{1-\sqrt{r/2\rk}} dx
\left(\sqrt{r\over 8\rk} + x\right)^{\gamma-1/2}.
\end{eqnarray}
\label{eq:denbound}
\end{subequations}
This density is plotted in Figure~\ref{fig:append}b for four values of
$\gamma$.
The integrals $I_1, I_2$ can be expressed in terms of hypergeometric
functions; the quantity $I_1(\gamma,0)$ that appears in the denominator
of equation~(\ref{eq:denbounda}) is
\beq
I_1\left(\gamma,0\right) = {\sqrt{\pi}\Gamma\left(\gamma+1/2\right)\over
\Gamma\left(\gamma+1\right)}.
\eeq

We stress that the density expressed by equations~(\ref{eq:denbound})
does not represent a steady-state distribution.
After phase-mixing, the density of stars will be non-zero at
all radii $0\le r\le\infty$ and the bound cloud will 
be elongated along the $x$-axis, 
as discussed above and as shown in
Figure~\ref{fig:frames}. 
However the fact that the density of bound stars is spherically
symmetric in configuration space {\it immediately
after} the kick allows the mass of the bound
population to be straightforwardly computed:
\begin{equation}
\mb = 4\pi\int_0^{8\rk} dr r^2 \rho_{\rm k}(r).
\end{equation}
Writing
\beq
2\mb = F_1(\gamma)M_{\rm k}, \ \ \ \ 
M_{\rm k} = 2\mh \left({G\mh\over \rh\vk^2}\right)^{3-\gamma}
\eeq
as above,
we have finally
\beq
F_1(\gamma) = {2\left(3-\gamma\right)\over I_1(\gamma,0)}
\int_0^8 dz\ z^{2-\gamma} \left[I_1(\gamma,z\rk) - 
\left(2z\right)^{(\gamma-1/2)/2}I_2(\gamma,z\rk)\right],
\ \ \ \ z = r/\rk.
\eeq
This function is plotted as the solid line in the top panel
of Figure~\ref{fig:fgamma}, where it is compared with the
approximate expression $11.6\gamma^{-1.75}$.

\appendix
\begin{center}
{\bf B. Glossary of acronyms and variables}\\
\end{center}

\noindent
BH: Black Hole \\
DGTO: Dwarf-Globular Transition Object \\
GC: Globular Cluster \\
GH: Gauss-Hermite (expansion) \\
GW: Gravitational Wave \\
IMF: Initial Mass Function \\
HCSS: HyperCompact Stellar System \\
SMBH: SuperMassive Black Hole \\
S/N: Signal-to-Noise ratio \\
UCD: UltraCompact Dwarf (galaxy) \\

\noindent
$\gamma$: power-law index for pre-kick density scaling with radius \\
$\Lambda$: Coulomb logarithm for 2-body relaxation \\
$\xi$: dimensionless radius in Dehnen profile; eqn.\ (\ref{eq:dehnen}) \\
$\rho(r)$: stellar density profile around BH \\
$\rho_{\rm 0}$: fiducial stellar density at $r_{\rm 0}$ \\
$\sigma$: 1-D velocity dispersion of the pre-kick galactic bulge \\
$\sigma_{\rm 0}$: width of Gaussian term in GH expansion \\
$\sigma_{\rm c}$: velocity dispersion for GH expansion \\
$\sigma_{\rm obs}$: observed velocity dispersion of the post-kick
HCSS; eqn.\ (\ref{eq:defs}) \\
$\Sigma(R)$: projected stellar surface density \\
$\phi(M)$: initial mass function; eqn.\ (\ref{eq:imf}) \\
$\Phi(M,z)$: distribution function of merging BHs with individual mass $M$
at redshift $z$; eqn.\ (\ref{eq:phimz}) \\
$\Omega_{\Lambda}$: cosmological density parameter for dark energy \\
$\Omega_{\rm m}$: cosmological density parameter for matter \\
$a$: semi-major axis of pre-merger BH binary orbit \\
$a_{\rm eq}$: semi-major axis of BH binary orbit at point when
dominated by GW losses\\
$a_{\rm h}$: semi-major axis of pre-merger {\it bound} BH binary
orbit; eqn.\ (\ref{eq:ah}) \\
$a_{\rm 1}$: spin parameter of larger pre-merger BH\\
$a_{\rm 2}$: spin parameter of smaller pre-merger BH\\
$d$: distance of passage from galactic center for collisionless
loss-cone \\
$d_{\rm sym}$: distance HCSS travels after kick before symmeterizing
into an elongated spheriod; eqn.\ (\ref{eq:dsym}) \\
$F_{\rm 1}(\gamma)$: dimensionless scaling function relating $M_{\rm
  k}$ and $M_{\rm b}$; eqns.\ (\ref{eq:F_1}, \ref{eq:fofgamma}) \\
$F_{\rm 2}(\gamma)$: dimensionless scaling function relating $r_{\rm
  eff}$ and $r_{\rm k}$; eqns.\ (\ref{eq:f2}, \ref{eq:reff}) \\
$F_{\rm 3}(\gamma)$: dimensionless scaling function relating $\sigma_{\rm
  obs}$ and $V_{\rm k}$; eqns.\ (\ref{eq:defs}, \ref{eq:sofgamma}) \\
$f_{\rm b}$: fraction of bound stellar mass $M_{\rm b}$ relative to BH
mass $M_{\rm BH}$; eqn.\ (\ref{eq:f1}) \\
$f(M)$: distribution function of merging BHs as a function of mass;
eqn.\ (\ref{eq:phimz}) \\
$G_{\rm m}(\gamma)$: dimensionless scaling function relating $M_{\rm
  b}$, $M_{\rm BH}$, and $r_{\rm eff}$ in collisionless regime; eqns.\
(\ref{eq:G1}, \ref{eq:G2}) \\
$G_{\rm s}(\gamma)$: dimensionless scaling function relating $M_{\rm
  b}$, $\sigma$, and $r_{\rm eff}$ in collisionless regime; eqns.\
(\ref{eq:G1}, \ref{eq:G2}) \\ 
$g(z)$: distribution function of merging BHs as a function of redshift;
eqn.\ (\ref{eq:phimz}) \\
$h$: dimensionless Hubble expansion parameter \\
$h_{\rm 4}$: measure of deviation from Gaussian in GH expansion \\
$H_{\rm m}(\gamma)$: dimensionless scaling function relating $M_{\rm
  b}$, $M_{\rm BH}$, and $\sigma_{\rm obs}$ in collisionless regime; eqns.\
(\ref{eq:H1}, \ref{eq:H2}) \\
$H_{\rm s}(\gamma)$: dimensionless scaling function relating $M_{\rm
  b}$, $\sigma$, and $\sigma_{\rm obs}$ in collisionless regime; eqns.\
(\ref{eq:H1}, \ref{eq:H2}) \\ 
$K(\gamma)$: dimensionless scaling function relating $M_{\rm
  b}$ and $M_{\rm BH}$; eqn.\ (\ref{eq:mvsr}) \\
$L$: bolometric luminosity of HCSS \\
$M_1$: mass of larger pre-merger BH \\
$M_2$: mass of smaller pre-merger BH \\
$M_{\rm b}$: post-kick mass in bound stars; eqn.\ (\ref{eq:f1}) \\
$M_{\rm BH}$: mass of the final, post-kick BH \\
$M_{\rm core}$: total mass in stars ejected from galactic core; eqn.\
(\ref{eq:m3}) \\
$M_{\rm D}$: total stellar mass in Dehnen (post-kick) density profile;
eqn.\ (\ref{eq:md}) \\
$M_{\rm k}$: pre-kick mass in stars within $r_{\rm k}$; eqn.\
(\ref{eq:mk1}) \\
$N$: multiplier of $\sigma$ which gives escape velocity from
core-Sersic galaxy; eqn. (\ref{eq:Nsigma}) \\
$\dot{N}$: post-kick rate of tidal disruptions; eqn.\ (\ref{eq:RR}) \\
$N(V)$: distribution function of line-of-site velocities \\
$q$: mass ratio of binary BH $M_2/M_1$ \\
$r$: radial distance from center of stellar cluster \\
$R$: projected radial distance from center of stellar cluster \\
$R_{\rm b}$: projected break radius for core-Sersic profile \\
$R(M,z)$: merger rate of binary BHs with total mass $M$ at redshift
$z$; eqn.\ (\ref{eq:rmz}) \\
$\rh$: pre-kick radius containing $2 M_{\rm BH}$ in stars \\
$r_{\rm 0}$: fiducial radius used to normalize pre-kick density
profile \\
$r_{\rm D}$: scaling radius for Dehnen density profile; eqn.\
(\ref{eq:dehnen}) \\
$r_{\rm eff}$: effective projected radius of post-kick stellar density
profile; eqn.\ (\ref{eq:f2}) \\
$r_{\rm infl}$: influence radius; eqn.\ (\ref{eq:defrinfl}) \\
$r_{\rm k}$: kick radius; eqn.\ (\ref{eq:rk}) \\
$r_{\rm t}$: tidal disruption radius; eqn.\ (\ref{eq:RR}) \\
$t_{\rm k}$: time elapsed since kick \\
$t_{\rm R}$: relaxation time of pre-merger stellar nucleus \\
$t_{\rm sym}$: time elapsed after kick before HCSS symmeterizes
into an elongated spheriod; eqn.\ (\ref{eq:tsym}) \\
$V_{\rm esc}$: escape velocity from host galaxy \\
$V_{\rm k}$: initial kick velocity of merged BH \\
$\mathbf{V}_{\rm k}$: 3-vector representation of kick velocity \\

\end{document}